\documentclass[a4paper,11pt]{article}
\usepackage{jcappub}
\usepackage{booktabs}

\newcommand{\ud}{{\rm d}}

\newcommand\numberthis{\addtocounter{equation}{1}\tag{\theequation}}

\title{Linear perturbation theory and structure formation in a Brans--Dicke theory of gravity without dark matter}

\author[a,b]{Lorenzo Gervani,}
\author[a,b,c]{Antonaldo Diaferio,}
\author[a,b,d]{Francesco Pace,}
\author[e]{Andrea P. Sanna}

\affiliation[a]{Dipartimento di Fisica, Universit\`a degli Studi di Torino, Via P. Giuria 1, I-10125, Torino, Italy}
\affiliation[b]{INFN - Sezione di Torino, Via P. Giuria 1, I-10125, Torino, Italy}
\affiliation[c]{Accademia delle Scienze di Torino, Via Maria Vittoria 3, I-10123, Torino, Italy}
\affiliation[d]{INAF - Istituto Nazionale di Astrofisica, Osservatorio Astrofisico di Torino, strada Osservatorio 20, 10025, Pino Torinese, Italy}
\affiliation[e]{INFN
Sezione di Roma, Piazzale Aldo Moro 5, 00185, Roma, Italy}

\emailAdd{lorenzo.gervani@unito.it}
\emailAdd{antonaldo.diaferio@unito.it}
\emailAdd{francesco.pace@unito.it}
\emailAdd{asanna@roma1.infn.it}

\abstract{We investigate the formation of the large-scale cosmic structure in a scalar-tensor theory of gravity belonging to the class of the Brans--Dicke theories.
The universe contains baryonic matter alone and neither dark matter nor dark energy. The two arbitrary functions of the scalar field characterizing the kinetic term and the self-interaction potential are set to $W(\varphi)=-1$ and $V(\varphi) = -\Xi \varphi$, respectively, with $\Xi$ a positive constant. In the weak-field limit, the theory reduces to Refracted Gravity, a non-relativistic theory whose modified Poisson equation contains the scalar field $\varphi$ that provides the gravitational boost required to describe the dynamics of galaxies and galaxy clusters without dark matter. In a flat, matter-dominated, homogeneous and isotropic universe the same scalar field $\varphi$ drives the accelerated expansion of the universe and describes the observed redshift evolution of the Hubble-Lema\^\i tre parameter $H(z)$.
However, in the equation of the growth factor of the linear perturbation theory, the form of $V(\varphi)$ makes the coefficient of the source of the gravitational field  proportional to $H^{-1}(z)$; therefore the gravitational field is strongly suppressed at early times and structure formation is delayed to redshift $z< 1$, in disagreement with the observation of formed galaxies at much larger redshifts. In addition, the form of $W(\varphi)$ and a linear $V(\varphi)$ imply that $\varphi$ generates twice the gravitational boost on massive particles than on photons, with possible observable consequences on the gravitational lensing phenomenon. It remains to be investigated whether different choices of $W(\varphi)$ and $V(\varphi)$, that can still make the theory reduce to Refracted Gravity in the weak-field limit, might alleviate these problems.
}


\begin{document}
\maketitle
\flushbottom
\allowdisplaybreaks

\section{Introduction}

In the currently widely accepted $\Lambda$ cold dark matter ($\Lambda$CDM) cosmological model, gravitational interactions are governed by General Relativity (GR) and the energy budget of the universe is dominated by the contributions of collisionless non-baryonic CDM and a positive cosmological constant $\Lambda$ that generates an accelerated expansion at the present time. CDM is invoked to address the missing mass problem on galactic scales~\cite{deMartino:2020gfi, Iocco:2015xga}, the clustering of the large-scale structure~\cite{Clowe:2006eq, DelPopolo:2013qba}, the abundances of light elements~\cite{Cyburt:2015mya, Fields:2019pfx}, and the amplitude of the temperature anisotropies in the cosmic microwave background radiation (CMB)~\cite{Planck:2018vyg, Planck:2019nip}. 

Although $\Lambda$CDM is very successful at describing a substantial number of observational data sets, several tensions have emerged both on large cosmic scales and on galactic scales~\cite{CosmoVerse:2025txj}. Most notably, the persisting tension on the Hubble-Lema\^\i tre parameter $H_0$, between the value measured with the distance ladder~\cite{Riess:2019cxk} and the one inferred from CMB measurements~\cite{Verde:2019ivm}, remains unsolved.
The observationally inferred $^7$Li abundance is approximately 3 times lower than the Big-Bang nucleosynthesis expectation~\cite{Mathews:2019hbi}. The value of the cosmological constant $\Lambda$ associated with the vacuum energy level in quantum field theory is $\sim 120$ orders of magnitude lower than expected~\cite{Weinberg:1988cp, Padilla:2015aaa}, suggesting that $\Lambda$, rather than being a constant, might be associated with a dynamical quantity named dark energy, responsible for the accelerated expansion of the universe~\cite{Riess1998}. This possibility is also suggested by the most recent DESI results~\cite{DESI:2025zgx}. 

N-body simulations of a CDM universe are also challenged by a number of unresolved problems on galactic scales, such as the core-cusp problem~\cite{McGaugh:1998tq},
the dwarf galaxy problem~\cite{Moore:1999nt, Klypin:1999uc}, the too-big-to-fail problem~\cite{Boylan-Kolchin:2011qkt} and the
plane of satellite problem~\cite{Pawlowski:2014una}. In addition, some relations, such as the radial acceleration (RAR)~\cite{McGaugh:2016leg, Lelli:2016cui} and the baryonic Tully-Fisher relation~\cite{McGaugh:2000sr}, are not naturally explained in the CDM paradigm, requiring a fine-tuning for the feedback of the baryonic processes to explain the emergence of the acceleration scale $a_0 = 1.2 \times 10^{-10} \text{m} \, \text{s}^{-2}$~\cite{deMartino:2020gfi}. 
Some of these problems may be partially solved, without abandoning CDM, by assuming halo models different from the Navarro-Frenk-White profile~\cite{Gentile:2004tb}.

An additional challenge is the fact that, although weakly interactive massive particles (WIMPs) have been considered the most likely candidate for CDM for decades, no definitive proof of their existence has yet been achieved. Dropping the collisionless assumptions, alternative dark matter candidates like warm dark matter, self-interacting dark matter, QCD axions, and fuzzy dark matter might solve some of the above-mentioned tensions~\cite{deMartino:2020gfi}, but they all still await clear observational confirmation (see for example~\cite{galanti2025}).
For more complete discussions on these and other challenges faced by the $\Lambda$CDM model we refer the reader to~\cite{Famaey:2011kh, Bullock:2017xww, deMartino:2020gfi}.

An alternative to the CDM paradigm is the suggestion that our current description of gravity must be modified. For example, since its proposal in the '80s, MOdified Newtonian Dynamics (MOND)~\cite{Milgrom:1983ca, Milgrom:1983pn, Milgrom:1983zz} has achieved notable success in addressing the phenomenological challenges of galactic dynamics—often making accurate predictions~\cite{Merritt2020}. However, formulating a covariant extension of MOND has proven extremely difficult~\cite{Famaey:2011kh, Skordis:2020eui}. This lack of a viable covariant framework has so far prevented MOND from making reliable predictions at cosmological scales, where a relativistic treatment is essential. 

A class of modified gravity theories attempts to explain both the dynamics of cosmic structure and the accelerated expansion of the universe by replacing both dark matter and dark energy with a single cosmic fluid, for instance a scalar field. These Unified Dark Matter (UDM) models~\cite{Bertacca:2008uf, Camera2009, Bertacca:2010ct, Camera2011} can provide a valid alternative to $\Lambda$CDM if the speed of sound associated with the clustering of this field is small enough that it can clump  and form cosmic structure. 

Here, we investigate a Brans--Dicke theory of gravity~\cite{Brans:1961sx} where a single scalar field still plays the role of both dark matter and dark energy. However, unlike the UDM models mentioned above, in this model, the scalar field does not need to cluster, because the role of the dark matter is played by a gravitational boost driven by the scalar field, rather than by its  clustering.

As was shown in~\cite{Sanna}, choosing the potential $V(\varphi)=-\Xi\varphi$ of the scalar field $\varphi$, with $\Xi$ a positive constant, allows to reconnect the weak-field limit of this Brans--Dicke theory with Refracted Gravity (RG). First introduced by~\cite{Matsakos}, RG is a classical theory of gravity that has been shown to reproduce the observed dynamics of galaxies and galaxy clusters~\cite{Cesare:2020pyz, Cesare:2022khd, Pizzuti}. This result is achieved by introducing a gravitational permittivity function, $\epsilon(\rho)$, in the standard Poisson equation, with $\rho$ the density of the baryonic matter. It turns out that in the weak-field limit, $\varphi=2\epsilon(\rho)$, thus linking $\varphi$ to the missing gravity problem on small scales. The transition between the Newtonian and the RG regimes happens at an acceleration scale $a_\Xi = (2\Xi - 8\pi G\rho/\varphi)^{1/2}$, that, being $\Xi \sim \Lambda$ as shown in~\cite{Sanna} and in the limit $2\Xi \gg 8\pi G\rho/\varphi$, returns a value comparable to the acceleration scale $a_0 \approx 10^{-10} \text{m s}^{-2}$ that emerges from the data on galaxy scales and is adopted by MOND~\cite{Li:2018tdo,Milgrom:2020cch}. 
Additionally, by using an effective dark energy approach, the scalar field $\varphi$ was shown to be responsible for the accelerated expansion of the universe~\cite{Sanna}. Therefore, in this theory, a single scalar field is able to account for both dark energy, on large scales, and dark matter on the scale of galaxies and galaxy clusters.
However, the growth of the large-scale structure in this theory remains to be explored. 

Here, we develop the linear perturbation theory to investigate whether the observed cosmic structure can  grow from small density perturbations in the early universe governed by this Brans--Dicke theory. Section \ref{Review of Refracted Gravity} introduces the main aspects of the Brans--Dicke theory and links it to RG, of which we also provide a short review. Section \ref{Homogenous and isotropic universe in CRG} provides analytical solutions for the background evolution of a flat and matter-dominated universe, and compares the prediction for the evolution of the Hubble-Lema\^\i tre parameter over time with current measures. Section \ref{CRG linear perturbation theory} introduces small perturbations on an isotropic and homogeneous background and develops a linear perturbation theory in the Brans--Dicke theory, deriving the growth factor equation and the sound speed of the scalar field for this model. We also discuss the Laplacians of the gravitational potentials of the perturbed metric, and their role in the gravitational lensing phenomenon. Section \ref{sec:QSA} uses the quasi-static approximation to provide a different route to pin down the role of the specific choices of the degrees of freedom of the Brans--Dicke theory in the evolution of the density perturbations.  Section \ref{Conclusions} draws conclusions and analyzes future prospects for this Brans--Dicke theory and for RG.

\section{Brans--Dicke theory and Refracted Gravity}
\label{Review of Refracted Gravity}
Scalar-tensor theories modify the standard Einstein--Hilbert action with the addition of a new scalar degree of freedom $\varphi$. Specifically, the generalized Brans--Dicke theory~\cite{Brans:1961sx, Quiros} has an action that, in the Jordan frame, reads\footnote{Throughout the paper, we adopt units with the light speed $c = 1$.}
\begin{equation}
    \mathcal{S} = \mathcal{S}_{\text{grav}} + \int d^{4} x \sqrt{-g} \mathcal{L}_{\mathrm{m}}\left(g_{\mu \nu}, \psi_{\mathrm{m}}\right)\,, 
    \label{CRG action}
\end{equation}
where\footnote{Sanna et al.~\cite{Sanna} use the same expression as in Eq.~(\ref{CRG action}), but follow the Weinberg convention for the sign of the Riemann tensor~\cite{Weinberg}. We choose instead to follow the opposite convention, to align more closely with the literature~\cite{Quiros, Kobayashi:2019, Tsujikawa}. We thus define $R_{\mu \nu} = \partial_{\alpha} \Gamma_{\mu \nu}^{\alpha} - \partial_{\nu} \Gamma_{\mu \alpha}^{\alpha} + \Gamma_{\mu \nu}^{\alpha} \Gamma_{\alpha \beta}^{\beta} - \Gamma_{\mu \beta}^{\alpha} \Gamma_{\alpha \nu}^{\beta}$. It follows that here the second and third terms of $\mathcal{S}_{\text{grav}}$, which involve the potentials $W(\varphi)$ and $V(\varphi)$, appear with the opposite sign with respect to~\cite{Sanna}, so that, overall, it results that our $\mathcal{S}_{\text{grav}}$ has the opposite sign of $\mathcal{S}_{\text{grav}}$ in~\cite{Sanna}. 
Consequently, the final equations of motion, which we will obtain through a variational principle below, actually coincide with those of~\cite{Sanna} when written explicitly in terms of the metric components, while they differ by some signs when these components are kept implicit in the Ricci tensor.}
\begin{equation}
   \mathcal{S}_{\text{grav}} = \frac{1}{16 \pi G} \int d^{4} x \sqrt{-g}\left[\varphi R - \frac{W(\varphi)}{\varphi} \nabla^{\alpha} \varphi \nabla_{\alpha} \varphi - 2V(\varphi)\right]\,.
   \label{explicit CRG action}
\end{equation}
Here, $g$ is the determinant of the metric tensor $g_{\mu\nu}$, $R = g^{\mu\nu}R_{\mu\nu}$ is the Ricci scalar, $\mathcal{L}_{\mathrm{m}}$ is the Lagrangian density associated with the matter fields $\psi_{\mathrm{m}}$, and $W(\varphi)$ and $V(\varphi)$ are arbitrary functions of the scalar field. In the limit $\varphi \xrightarrow[]{} \bar \varphi = \text{constant}$, the action reduces to the Einstein--Hilbert action with a cosmological constant $V(\bar \varphi)$. 
One relevant aspect of this action is the fact that, in the absence of matter sources, it is invariant under a simultaneous rescaling of the scalar field and the metric. It follows that the present value of the scalar field, which we call $\varphi_0$, can be chosen arbitrarily in the vacuum solution, and carries no physical meaning. On the contrary, if the matter term is present, the value of $\varphi_0$ determines the coupling strength between matter and gravity through the effective Newton constant, as it will be shown more clearly below. Thus, $\varphi_0$ will be treated as one of the fundamental parameters of the theory. 

The action in Eq.~(\ref{explicit CRG action}) belongs to the broader family of Horndeski theories~\cite{Horndeski:1974wa, Kobayashi:2011nu}, which represent the most general scalar-tensor theories that still admit second order equations of motion. The Lagrangian density of the Horndeski theories is usually written as $\mathcal{L} = \sum_{i = 2}^5 \mathcal{L}_i $, where each of the four terms is related to a scalar function, $G_i\left(\varphi, X\right)$, as follows:
\begin{equation}
    \begin{array}{l}
\mathcal{L}_{2} \equiv G_{2}(\varphi, X)\,, \\
\mathcal{L}_{3} \equiv G_{3}(\varphi, X) \square \varphi\,, \\
\mathcal{L}_{4} \equiv G_{4}(\varphi, X) R-2 G_{4, X}(\varphi, X)\left[(\square \varphi)^{2}-\left(\nabla_{\mu} \nabla_{\nu} \varphi\right)^{2}\right]\,, \\
\mathcal{L}_{5} \equiv G_{5}(\varphi, X) G_{\mu \nu} \nabla^{\mu}\nabla^\nu \varphi+\cfrac{1}{3} G_{5, X}(\varphi, X)\left[(\square \varphi)^{3}-3 \square \varphi\left(\nabla_{\mu} \nabla_{\nu} \varphi\right)^{2}+2\left(\nabla_{\mu} \nabla_{\nu} \varphi\right)^{3}\right]\,,
\label{Horndeski}
\end{array}
\end{equation}
where $X = g^{\mu\nu}\nabla_\mu\varphi\nabla_\nu\varphi$ is the canonical kinetic term of the scalar field. In Eqs.~(\ref{explicit CRG action}) and~(\ref{Horndeski}) we take the scalar field to be a dimensionless quantity, written in units of the Planck mass $M_{\text{pl}}= \left(8\pi G\right)^{-1/2}$. An alternative formulation, often used in the literature, is to adopt $\phi = M_{\text{pl}}\varphi$. The expressions in Eq.~(\ref{Horndeski}) imply that the Horndeski functions of the Brans--Dicke theory are
\begin{equation}
\begin{aligned}
   G_4\left(\varphi, X\right) & = \frac{M_{\text{pl}}^2}{2}\varphi\,; \hspace{5mm} G_2\left(\varphi, X\right) = \frac{M_{\text{pl}}^2}{2}\left( - \frac{W(\varphi)}{\varphi} \nabla^{\alpha} \varphi \nabla_{\alpha} \varphi - 2V(\varphi) \right)\,;\\
   G_3\left(\varphi, X\right) & = G_5\left(\varphi, X\right) = 0 \,.
   \end{aligned}
   \label{Horndeski functions of CRG}
\end{equation}
By varying Eq.~(\ref{CRG action}) with respect to the metric, we find~\cite{Quiros}
\begin{equation}
    \varphi\left(R_{\mu \nu}-\frac{1}{2} g_{\mu \nu} R\right) + \left(\frac{W}{2 \varphi} \nabla^{\alpha} \varphi \nabla_{\alpha} \varphi+\square \varphi+V\right) 
    g_{\mu \nu} - \nabla_{\mu} \nabla_{\nu} \varphi - \frac{W}{\varphi} \nabla_{\mu} \varphi \nabla_{\nu} \varphi = 8 \pi G T_{\mu \nu}\,,
    \label{Modified Einstein}
\end{equation}
where $T_{\mu \nu}\sqrt{-g} = -2\delta\left(\sqrt{-g}\mathcal{L}_{\mathrm{m}}\right)/\delta g^{\mu\nu}$.
The variation with respect to the scalar field yields instead
\begin{equation}
    \varphi R - \left(\frac{W}{\varphi}-\frac{d W}{d \varphi}\right) \nabla^{\alpha} \varphi \nabla_{\alpha} \varphi + 2W \square \varphi - 2 \varphi \frac{d V}{d \varphi}=0\,.
    \label{Modified KG}
\end{equation}
In Eqs.~(\ref{Modified Einstein}) and~(\ref{Modified KG}) the $\varphi$ dependence of $W$ and $V$ has been left implicit. 

As was first shown in~\cite{Sanna}, the specific theory defined by the two choices
\begin{equation}
    W(\varphi) \equiv -1; \hspace{3mm} V(\varphi) = -\Xi\varphi\,, \quad \Xi = \text{constant} \,,
    \label{choices}
\end{equation}
(where $\Xi$ is taken to be positive) reduces to Refracted Gravity (RG) in the weak-field limit. RG, first introduced by~\cite{Matsakos}, is a non-relativistic theory of gravity; it is inspired by an analogy to electromagnetism, in particular by the well-known fact that in a dielectric medium of non-uniform permittivity, both the direction and the magnitude of the electromagnetic field are affected. Postulating that gravity possesses a similar ``gravitational permittivity'' -- assumed to be a monotonically increasing function of the local mass density $\rho$ -- the non-relativistic formulation of RG was shown to be able to describe a number of observed features of the dynamics of galaxies and galaxy cluster~\cite{Matsakos, Cesare:2020pyz, Cesare:2022khd, Pizzuti}. 
The dynamics of RG is ruled by a modified version of the Poisson equation
\begin{equation}
    \nabla \cdot \left(\epsilon \nabla \Psi\right) = 4\pi G \rho \,,
    \label{RG equation}
\end{equation}
where $\Psi$ is the gravitational potential and $G$ is the gravitational constant.
Contrary to electrodynamics, the gravitational permittivity $\epsilon = \epsilon\left(\rho\right)$ takes values $0 < \epsilon_0 < \epsilon\left(\rho\right) < 1$, where $\epsilon_0 = \epsilon\left(0\right)$ is the permittivity of the vacuum. 
Since Newtonian gravity accurately describes stellar dynamics, the standard Poisson equation must be recovered from Eq.~(\ref{RG equation}) above a certain threshold density. Deviations from the Newtonian predictions are instead observed in the outskirts of galaxies and on larger scales where the density is expected to be smaller than this threshold. So, one can assume a permittivity of the form
\begin{equation}
    \epsilon(\rho) \simeq \left\{ \begin{aligned}
        1 \hspace{3mm} \text{for} \hspace{3mm} \rho \gtrsim \rho_{\text{thr}}\,, \\
        \epsilon_0 \hspace{3mm} \text{for} \hspace{3mm} \rho \lesssim \rho_{\text{thr}}\,.
    \end{aligned} \right.
\end{equation}
Eq.~(\ref{RG equation}) is structurally similar to the MOND equation~\cite{Bekenstein}, where, however, the function $\epsilon$ depends on the local acceleration, $\epsilon = \epsilon(\nabla\Psi)$, and not on the local mass density $\rho$.

With Eq.~(\ref{choices}), the equations of motion are
\begin{align*}
     & \varphi\left(R_{\mu \nu}-\frac{1}{2} g_{\mu \nu} R\right) + \left(-\frac{1}{2 \varphi} \nabla^{\alpha} \varphi \nabla_{\alpha} \varphi+\square \varphi-\Xi \varphi\right) g_{\mu \nu} \\
    &\quad - \nabla_{\mu} \nabla_{\nu} \varphi + \frac{1}{\varphi} \nabla_{\mu} \varphi \nabla_{\nu} \varphi = 8 \pi G T_{\mu \nu}\,, \numberthis
    \label{Modified Einstein 2}
\end{align*}
\begin{equation}
    \varphi R+\frac{1}{\varphi} \nabla^{\alpha} \varphi \nabla_{\alpha} \varphi - 2 \square \varphi+2 \Xi \varphi=0\,.
    \label{Modified KG 2}
\end{equation}
We can now show that the $00$-component of the modified Einstein equations, Eq.~(\ref{Modified Einstein 2}), reduces to its non-relativistic counterpart, Eq.~(\ref{RG equation}), in the weak-field limit. By using Eq.~(\ref{Modified KG 2}) into Eq.~(\ref{Modified Einstein 2}), we obtain
\begin{equation}
    \varphi R_{\mu \nu} - \nabla_{\mu} \nabla_{\nu} \varphi + \frac{1}{\varphi} \nabla_{\mu} \varphi \nabla_{\nu} \varphi = 8 \pi G T_{\mu \nu}\, .
    \label{Rearrenged Modified Einsten}
\end{equation}
Contracting this result with $g^{\mu\nu}$ and substituting into Eq.~(\ref{Modified KG 2}) yields
\begin{equation}
    \square \varphi - 2\Xi\varphi = 8\pi G T\,,
    \label{Rearrenged KG}
\end{equation}
where $T$ is the trace of the stress-energy tensor.
By expanding around the Minkowski metric,
\begin{equation}
    g_{\mu \nu} \simeq \eta_{\mu \nu}+h_{\mu \nu}\,, \\
\end{equation}
where
\begin{equation}
    g_{00} \simeq \eta_{00} - 2\Psi\,, \hspace{3mm}
g_{0 i} \simeq 0\,, \hspace{3mm} 
g_{i j} \simeq \eta_{ij} - 2\Phi\delta_{ij}\,,
\end{equation}
with $\Psi$ and $\Phi$ the two gravitational potentials, and by computing the corresponding Christoffel symbols, one finds that Eq.~(\ref{Rearrenged Modified Einsten}) becomes, up to first order in perturbatively small quantities $\Phi$ and $\Psi$,
\begin{equation}
    \nabla \cdot \left(\varphi \nabla \Psi\right) = 8\pi G \rho \,.
\end{equation}
Thus, the classical RG equation is recovered if one identifies the scalar field with twice the permittivity, $\varphi = 2\epsilon$. The Newtonian regime is therefore recovered for $\varphi = 2$.

We note that the weak-field limit of the modified Einstein equation, Eq.~(\ref{Modified Einstein}), for a Brans--Dicke theory of action Eq.~(\ref{explicit CRG action}) with $W(\varphi) = -1$ and a generic potential $V(\varphi)$ is
\begin{equation}
    \varphi \nabla^2\Phi + \nabla\Phi\nabla\varphi - (1+2\Phi)\left(V(\varphi) - \varphi\frac{{\rm d}V}{{\rm d}\varphi}\right) = 8\pi G \rho\,.
\end{equation}
It follows that the condition we need to impose on the potential $V(\varphi)$ to obtain the RG Poisson equation, Eq.~(\ref{RG equation}), as the weak-field limit is 
\begin{equation}
    V(\varphi) - \varphi\frac{{\ud} V}{{\ud}\varphi} = 0\,.
\end{equation}
Therefore, in Eq.~(\ref{choices}), the constant $\Xi$ can be taken either with a positive or negative sign. In the following analysis, we chose the positive sign, at least initially, to compare our results with those of \cite{Sanna}.

We end this section by remarking that there are actually two distinct interpretations of the modifications induced by the coupling between gravity and the scalar field. They can be interpreted as either a modification of the spacetime geometry, as implied by Eq.~(\ref{Modified Einstein 2}), or as an effective source term in the field equations. Following this latter interpretation, the modified Einstein equations can also be recast as~\cite{Quiros}
\begin{equation}
     \varphi G_{\mu\nu} = 8\pi G \left(T_{\mu\nu} + T_{\mu\nu}^\varphi\right)\,,
\end{equation}
where we have defined the scalar field stress-energy tensor
\begin{equation}
    T_{\mu\nu}^\varphi = \frac{1}{8 \pi G}\left[- \left(-\frac{1}{2 \varphi} \nabla^{\alpha} \varphi \nabla_{\alpha} \varphi+\square \varphi-\Xi \varphi\right) g_{\mu \nu} + \nabla_{\mu} \nabla_{\nu} \varphi - \frac{1}{\varphi} \nabla_{\mu} \varphi \nabla_{\nu} \varphi \right]\,.
    \label{scalar field stress energy}
\end{equation}

\section{Homogeneous and isotropic universe}
\label{Homogenous and isotropic universe in CRG}
To explore the cosmological implications and predictions of the Brans--Dicke theory, we derive the modified Friedmann equations for a homogeneous and isotropic universe, whose geometry is described by the Friedmann-Lemaître-Robertson-Walker (FLRW) metric
\begin{equation}
    \mathrm{d} s^{2}=-\mathrm{d} t^{2}+a(t)^{2}\left(\frac{\mathrm{d} r^{2}}{1-k r^{2}}+r^{2} \mathrm{d} \theta^{2}+r^{2} \sin ^{2} \theta \mathrm{d} \phi^{2}\right)\,,
    \label{FRW}
\end{equation}
where $a(t)$ is the scale factor, $k$ is the spatial curvature parameter, $t$ is the cosmic time and $(r,\theta, \phi)$ are the spatial coordinates in the comoving coordinate system. In the expressions of this section, the scalar field is assumed to be a function of time only, and the dot sign refers to the derivation with respect to $t$. 
Using the explicit expressions for the Ricci tensor components, reported in Appendix~\ref{AppA}, the (00) component of Eq.~(\ref{Modified Einstein 2}) becomes
\begin{equation}
    H^2 = \frac{8\pi G}{3\varphi}\left(\rho+\rho_\varphi\right) - \frac{k}{a^2}\,,
    \label{modified Friedmann 1}
\end{equation}
where $H = \dot{a}/a$, $\rho$ is the matter and radiation density, and 
\begin{equation}
    \rho_\varphi = -\frac{1}{8\pi G}\left( \frac{\dot{\varphi}^2}{2\varphi} + 3H\dot{\varphi} +\Xi\varphi\right)\,,
\end{equation}
is the effective energy density associated with the scalar field. The (rr) component yields 
\begin{equation}
    \frac{\ddot{a}}{a} = -\frac{4\pi G}{3\varphi}\left(\rho + 3P + \rho_\varphi + 3P_\varphi\right)\,,
    \label{modified Friedmann 2}
\end{equation}
where $P$ is the matter and radiation pressure, and
\begin{equation}
    P_\varphi = -\frac{1}{8\pi G}\left(\frac{\dot{\varphi}^2}{2\varphi} - \ddot{\varphi} - 2H\dot{\varphi} - \Xi\varphi\right)\,,
\end{equation}
is the effective pressure of the scalar field. The expressions for the energy density and pressure of the scalar field can also be derived as the (00) and (ii) components, respectively, of the stress-energy tensor [Eq.~(\ref{scalar field stress energy})]. Eqs.~(\ref{modified Friedmann 1}) and~(\ref{modified Friedmann 2}) are the modified version of the Friedmann equations. Finally, using Eqs.~(\ref{R components}),~(\ref{Phi derivatives}) and the modified Friedmann equations, Eq.~(\ref{Modified KG 2}) yields
\begin{equation}
 \ddot{\varphi} + 3H\dot{\varphi} + 2\Xi\varphi = 8\pi G\left(\rho - 3P\right)\,,
    \label{modified KG}
\end{equation}
which is the Klein-Gordon (KG) equation.

\subsection{An analytic solution: flat, matter-dominated universe}

The coupled system of Eqs.~(\ref{modified Friedmann 1}),~(\ref{modified Friedmann 2}) and~(\ref{modified KG}) can be solved analytically in the special case of a spatially flat ($k=0$) and matter-dominated universe. By ``matter-dominated'' we mean that the stress-energy tensor $T_{\mu\nu}$ is dominated by the pressureless matter, while the radiation is neglected. Thus, the matter is assumed to follow a barotropic equation of state $P_\text{m} = w_{\text{m}} \rho_\text{m}$ with $w_{\text{m}}=0$. According to the continuity equation,\footnote{In the Jordan frame, in which we work, the stress-energy tensor of matter is covariantly-conserved, $\nabla_\mu T^{\mu\nu} = 0$~\cite{Faraoni:2004pi, Postma:2014vaa, Clifton:2011jh}. } the matter energy density scales like a power law, $\rho_\text{m}(t) = \rho_{\text{m}0}\, a(t)^{-3}$. By contrast, radiation follows the power law $\rho_\text{r}(t) = \rho_{\text{r}0}\, a(t)^{-4}$. Here, $\rho_{\text{m}0}$ and $\rho_{\text{r}0}$ are the energy densities of matter and radiation at the present time, respectively. In the following, we review the derivation performed in~\cite{Sanna}.

By dividing the three Eqs.~(\ref{modified Friedmann 1}), (\ref{modified Friedmann 2}) and (\ref{modified KG}) by $H^2$ and then using again Eq.~(\ref{modified KG}) to eliminate the terms proportional to $\Xi$ in Eqs.~(\ref{modified Friedmann 1}) and~(\ref{modified Friedmann 2}), we obtain
\begin{equation}
\begin{aligned}
\frac{1}{H^{2}} \frac{\ddot{a}}{a}-\frac{1}{3 H^{2}}\left(\frac{\dot{\varphi}^{2}}{\varphi^{2}}-\frac{\ddot{\varphi}}{\varphi}\right) & =-\frac{8 \pi G}{3 H^{2} \varphi} \rho_{\mathrm{m}}\,, \\
2+\frac{1}{H} \frac{\dot{\varphi}}{\varphi}+\frac{1}{3 H^{2}}\left(\frac{\dot{\varphi}^{2}}{\varphi^{2}}-\frac{\ddot{\varphi}}{\varphi}\right) & =\frac{8 \pi G}{3 H^{2} \varphi} \rho_{\mathrm{m}}\,, \\
\frac{1}{3 H^{2}} \frac{\ddot{\varphi}}{\varphi}+\frac{1}{H} \frac{\dot{\varphi}}{\varphi}+\frac{2}{3 H^{2}} \Xi & =\frac{8 \pi G}{3 H^{2} \varphi} \rho_{\mathrm{m}}\,.
\end{aligned}
\end{equation}
It is convenient to introduce the cosmological parameters
\begin{equation}
    \Omega \equiv \frac{2 \Omega_{\mathrm{m}}}{\varphi} \equiv \frac{16 \pi G}{3 H^{2} \varphi} \rho_{\mathrm{m}}\, , \qquad \Omega_{\Xi} \equiv \frac{\Xi}{3 H^{2}} \,,
    \label{omega, omega_xi}
\end{equation}
along with the combinations
\begin{equation}
    q \equiv-\frac{a \ddot{a}}{\dot{a}^{2}}=-\frac{\ddot{a}}{a H^{2}}\, , \qquad q_{\varphi} \equiv-\frac{\ddot{\varphi}}{\varphi H^{2}} \, , \qquad \zeta \equiv \frac{\dot{\varphi}}{H \varphi}\,. 
\end{equation}
The system of equations simplifies to
\begin{equation}
    \begin{aligned}
q+\frac{1}{3} \zeta^{2}+\frac{1}{3} q_{\varphi} & =\frac{1}{2} \Omega\,, \\
2+\zeta+\frac{1}{3} \zeta^{2}+\frac{1}{3} q_{\varphi} & =\frac{1}{2} \Omega\,, \\
-\frac{1}{3} q_{\varphi}+\zeta & =\frac{1}{2} \Omega-2 \Omega_{\Xi} \,.
\label{passaggio 1}
\end{aligned}
\end{equation}

The parameter $\Omega$ accounts for the matter density contribution. It is rescaled by a factor $2/\varphi$ with respect to its $\Lambda$CDM counterpart. Therefore, the contribution of the dark matter of the $\Lambda$CDM model to the total matter density parameter is replaced here by a boosted baryonic matter density. The effect is analogous to the gravity boost described by Eq.~(\ref{RG equation}) in the weak-field limit and on small scales, in line with the association of $\varphi$ to the gravitational permittivity $\epsilon$. The parameter $\Omega_\Xi$, instead, is the analogue of the $\Lambda$CDM parameter $\Omega_\Lambda$, so $\Xi$ has the role of a cosmological constant.

By combining the first two equations of Eqs.~(\ref{passaggio 1}) above yields $q - \zeta = 2$, which implies
\begin{equation}
    \frac{d}{d t} \ln \varphi=-\frac{d}{d t}(\ln \dot{a}+2 \ln a) \implies
    \varphi=\frac{H_{0} \varphi_{0}}{H a^{3}}\,,
    \label{phi analitico}
\end{equation}
where $H_0$ and $\varphi_0$ are the values of the Hubble-Lema\^\i tre parameter and the scalar field at the present time. By summing the first and the last equations in Eqs.~(\ref{passaggio 1}), instead, we obtain $q=\Omega-2 \Omega_{\Xi}-\zeta-\zeta^{2}/3$. Substituting Eq.~(\ref{phi analitico}) into this relation and the two definitions of Eqs.~(\ref{omega, omega_xi}) yields~\cite{Sanna}
\begin{equation}
    \left(\frac{d}{d t} \ln \dot{a}-\frac{d}{d t} \ln a\right)^{2}=\left(\frac{d}{d t} \ln \frac{\dot{a}}{a}\right)^{2} \implies \frac{d H}{d t}= \pm \sqrt{3}\left(H^{4}+\Omega_{0} H_{0} H^{3}-2 \Omega_{\Xi 0} H_{0}^{2} H^{2}\right)^{1 / 2}\,,
    \label{dH/dt}
\end{equation}
where $\Omega_0$ and $\Omega_{\Xi 0}$ are the cosmological parameters at the present time. 
The upper and lower signs in Eq.~(\ref{dH/dt}) can be interpreted as two different cosmologies, called CRG+ and CRG- respectively, where CRG stands for Covariant RG~\cite{Sanna}. To be consistent with observations, the Hubble-Lema\^\i tre parameter must necessarily decrease with increasing cosmic time, so we will later select the minus sign when comparing our model with the data. 

We now seek an expression for the Hubble-Lema\^\i tre parameter in terms of redshift, $H(z)$, or, equivalently, in terms of the scale factor, $H(a)$. This expression can be compared with available observational data to constrain the parameters of the model and assess whether our theory adequately accounts for the universe's background evolution. To this end, it is useful to recast Eq.~(\ref{dH/dt}) using $a$ as an independent variable instead of $t$. We obtain
\begin{equation}
    \frac{d H}{d a}= \pm \frac{\sqrt{3}}{a}\left(H^{2}+\Omega_{0} H_{0} H-2 \Omega_{\Xi 0} H_{0}^{2}\right)^{1 / 2}\,,
\end{equation}
which can be directly integrated to
\begin{equation}
    H(a) = -\frac{1}{2}\Omega_0 H_0 + \sqrt{\frac{1}{4}\Omega_0^2 + 2\Omega_{\Xi 0}}\frac{H_0}{2}\left[ \left(C + \sqrt{C^2 -1}\right)a^{\pm\sqrt{3}} + \left(C + \sqrt{C^2 -1}\right)^{-1}a^{\mp\sqrt{3}} \right]\,,
    \label{H(a)}
\end{equation}
where
\begin{equation}
    C \equiv \frac{1 + \frac{\Omega_0}{2}}{\sqrt{\frac{1}{4}\Omega_0^2 + 2\Omega_{\Xi 0}}} \geq 1\,.
\end{equation}
The inequality must be satisfied to guarantee that $H(a)$ is a real number. This condition can be recast as
\begin{equation}
    \Omega_0 > 0 \, , \qquad 0 < \Omega_{\Xi 0} \leq \frac{\Omega_0 + 1}{2}\,.
    \label{omega_xi condition}
\end{equation}

\subsection{Numerical background solution and \texorpdfstring{$H(z)$}{H} fit}
\label{Numerical background solution and $H(z)$ fit}
We now solve the system of Eqs.~(\ref{modified Friedmann 1}), (\ref{modified Friedmann 2}) and~(\ref{modified KG}) 
when the radiation contribution is also taken into account. We solve the system numerically. 
We thus check our analytical result for $H(a)$ and $\varphi(a)$ derived above and obtain some useful relations that will be employed for the numerical solutions of the perturbations evolution in Section \ref{CRG linear perturbation theory}. We adopt $N \equiv \ln a$ as our independent variable of integration. 

By considering that $\dot{\varphi} = aH {\rm d}\varphi/{\rm d} a = H {\rm d}\varphi/{\rm d}\ln a \equiv H\varphi'$, where the prime denotes differentiation with respect to $N$, Eq.~(\ref{modified Friedmann 1}) becomes
\begin{equation}
     H = \sqrt{\frac{\cfrac{8\pi G}{3\varphi}\rho - \cfrac{\Xi}{3}}{1 + \cfrac{\varphi^\prime}{\varphi} + \cfrac{\left(\varphi^\prime\right)^2}{6\varphi^2}}}\,,
     \label{H}
\end{equation}
whereas Eq.~(\ref{modified KG}) becomes
\begin{equation}
     \varphi^{\prime\prime} + \left(3 + \frac{H'}{H}\right)\varphi' + \frac{1}{H^2}\left[2\Xi\varphi -8\pi G\left(\rho - 3P\right)\right] = 0\,.
    \label{ODE}
\end{equation}
Solving Eq.~(\ref{ODE}) relies on $H'$. The latter can be derived by  differentiating Eq.~(\ref{H}) and using Eq.~(\ref{ODE}) to eliminate $\varphi^{\prime\prime}$ in the resulting expression. We obtain
\begin{align*}
     H' = \frac{1}{H\left(2+\frac{\varphi^\prime}{\varphi}\right)}\Bigg\{ & H^2\left[ 3\frac{\varphi^\prime}{\varphi} + 2\left(\frac{\varphi^\prime}{\varphi} \right)^2 + \frac{1}{3}\left(\frac{\varphi^\prime}{\varphi} \right)^3\right] + \left(1 + \frac{\varphi^\prime}{3\varphi} \right)\left(2\Xi -\frac{8\pi G}{\varphi}\rho_{\text{m}0}e^{-3N}\right) \\
    &  - \frac{8\pi G}{3\varphi^2}\varphi^{\prime}\left( \rho_{\text{m}0}e^{-3N} + \rho_{\text{r}0}e^{-4N} \right) + \frac{8\pi G}{3\varphi}\left( -3\rho_{\text{m}0}e^{-3N} - 4\rho_{\text{r}0}e^{-4N} \right) \Bigg\}\,. \numberthis
    \label{Hprime}
\end{align*}
Here, we have explicitly used the functional forms for $\rho_{\rm m}$ and $\rho_{\rm r}$
\begin{align}
    \rho(N) & = \rho_{\text{m}0}e^{-3N} + \rho_{\text{r}0}e^{-4N}\,, \\
    \left(\rho - 3P\right)(N) & = \rho_{\rm m} = \rho_{\text{m}0}e^{-3N}\,. \label{rho-3P}
\end{align}
The parameters $\rho_{\text{m}0}$, $\rho_{\text{r}0}$ and $H_0$ are considered known here and, for the sake of simplicity, are taken to be equal to the current values derived from the data. Indeed, the estimations of the baryonic density $\rho_{\text{m}0}$ from the data assume an underlying cosmological model, usually the $\Lambda$CDM model. Therefore, in principle, we would need to estimate this parameter from the data by assuming the Brans--Dicke cosmological model. The impact of this simplifying assumption on the comparison between the model and the values $H(z)$ derived from the data remains to be estimated. Notice that $\rho_{\text{m}0}$ and $\Omega_{\text{m}0}$ refer to the baryonic component only, so we adopt $\Omega_{\text{m}0} \sim 0.05$~\cite{Planck:2018vyg}. 

By contrast, the parameters $\Omega_{\Xi 0}$ and $\varphi_0$ are unknown, and can potentially be estimated by comparing the model to the observational data. $\Omega_{\Xi 0}$ and $\varphi_0$ are the only additional parameters of our model, as $H_0, \, \rho_{\text{m}0}, \, \rho_{\text{r}0}$ are also parameters of the $\Lambda$CDM model. The remaining unknown quantities, such as $\Xi$ and $\Omega_0$, depend on these parameters through the definitions of Eqs.~(\ref{omega, omega_xi}).

Once the pair $(\Omega_{\Xi 0}, \varphi_0)$ is fixed, Eq.~(\ref{ODE}), together with Eqs.~(\ref{H}) and~(\ref{Hprime}), can be numerically integrated to find the background evolution of the scalar field and of the Hubble-Lema\^\i tre parameter, through Eq.~(\ref{H}). The boundary conditions are fixed at the present time $t = t_0$, with $a(t_0) = a_0 = 1$ and $N = 0$. The initial conditions for the scalar field can be inferred from the analytical solution of Eq.~(\ref{phi analitico}), since radiation is negligible at the present time, 
\begin{equation}
    \varphi(N=0) = \varphi_0; \hspace{5mm} \varphi'(N=0) = -\varphi_0\left(\frac{\dot{H}_0}{H^2_0} + 3\right) \,,
\end{equation}
where $\dot{H}_0 = \dot{H}(t=t_0)$ according to Eq.~(\ref{dH/dt}). 

For the numerical integration, we employed the Runge-Kutta Cash-Karp method with an adaptive step size, which calculates fourth and fifth order accurate solutions and estimates the error as the difference between the two solutions.
\begin{figure}[!t]
         \includegraphics[width=0.5\textwidth]{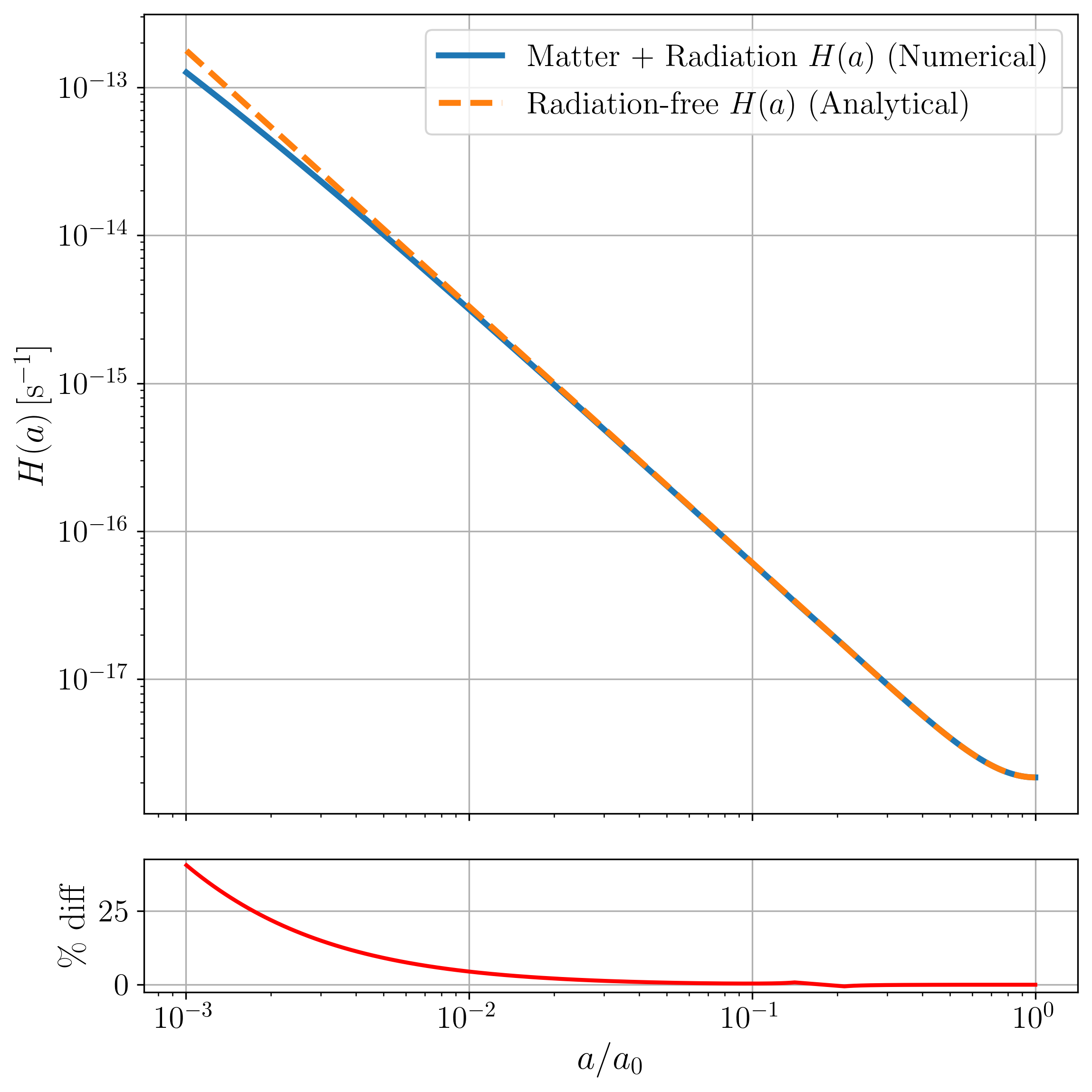}
        \includegraphics[width=0.5\textwidth]{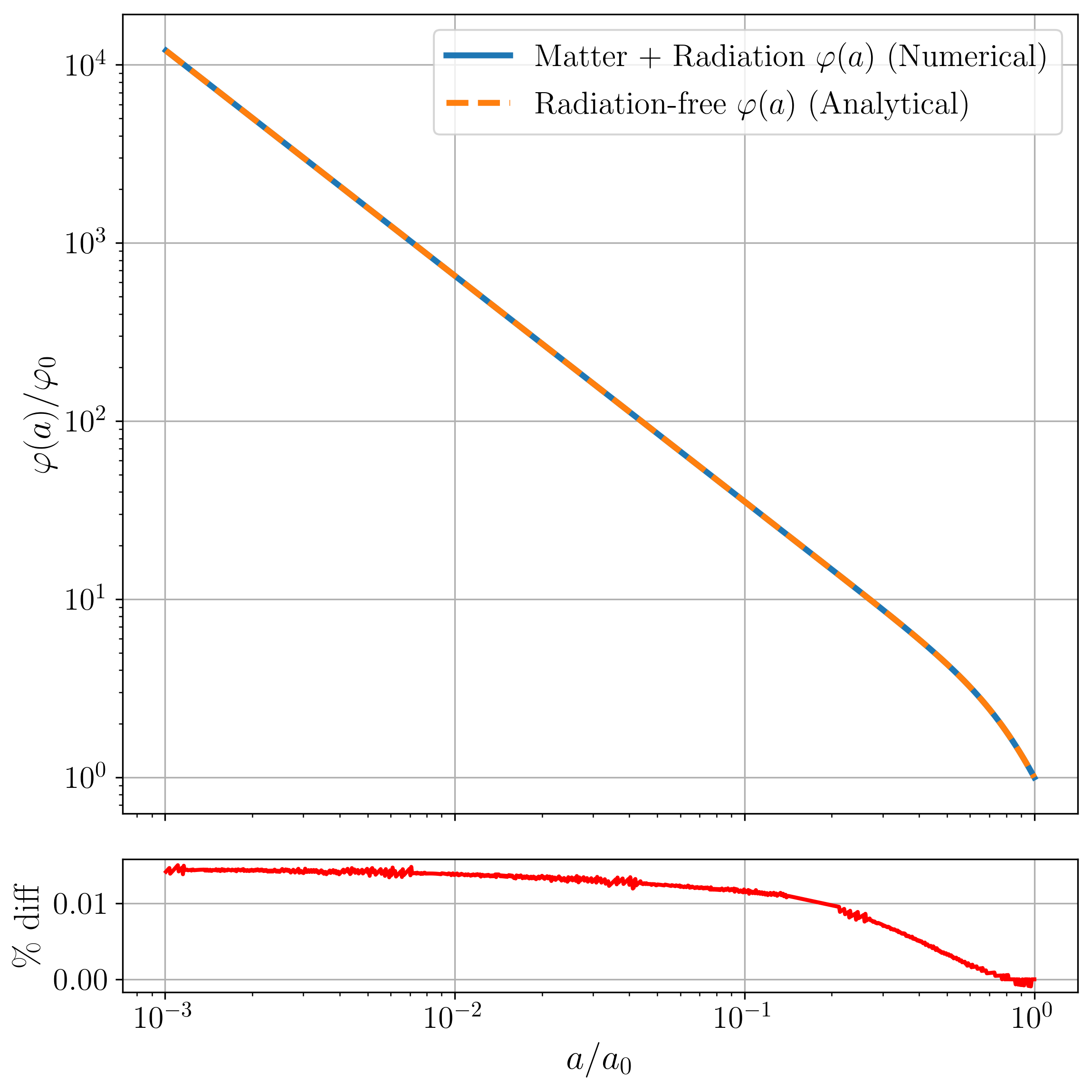}
\caption{Comparison between the evolution of $H(a)$ (left) and $\varphi(a)/\varphi_0$ (right) derived analytically in a radiation-free universe and numerically in a universe containing both matter and radiation. The bottom panels show the percentage difference between the analytical and numerical solutions. The two solutions agree with each other on the entire redshit range except for $H(a)$ at $z \gtrsim  100$.}
    \label{comparison analytical numerical}
\end{figure}
As shown in Figure \ref{comparison analytical numerical}, there is good agreement between the analytical and numerical solutions for both $H(a)$ and $\varphi(a)$ at small redshifts. However, for scale factors below $10^{-2}$, an increasing discrepancy appears in $H(a)$, due to the breakdown of the radiation-free approximation used to derive the analytical solution. In contrast, the numerical and analytical results for $\varphi(a)$ remain in good agreement across the entire range of $a$ considered. This consistency arises from the fact that only the combination $\rho - 3P$ enters the KG Eq.~(\ref{ODE}). As seen from Eq.~(\ref{rho-3P}), this combination only depends on the matter content, thus remaining consistent with the assumptions made in the analytical derivation.

Next, we constrain the model parameters $(\Omega_{\Xi 0}, \varphi_0)$ using observed data points for the Hubble-Lema\^\i tre parameter as a function of redshift. Two main data sets are available in the literature~\cite{Hu:2025fsz, Hu:2024big, CosmoVerse:2025txj}:  the cosmic chronometer data set~\cite{Moresco:2022phi, Moresco:2023zys}, and the baryonic acoustic oscillations (BAO) dataset~\cite{Yang:2024kdo}. For this analysis, we refer to the values reported in~\cite{Yu:2017iju} and only use data points based on cosmic chronometers. 
Indeed, in the cosmic chronometer approach, the estimate of the age of a galaxy is based on evolution models of stellar populations~\cite{Yu:2017iju, M_Moresco_2012, Zhang_2014} that do not assume any specific cosmological model. On the contrary, the BAO scale is computed by solving the Boltzmann equations in a chosen cosmological model. Estimates based on the BAO scale, thus, are model dependent. 

\begin{figure}[!t]
        \includegraphics[width=0.5\textwidth]{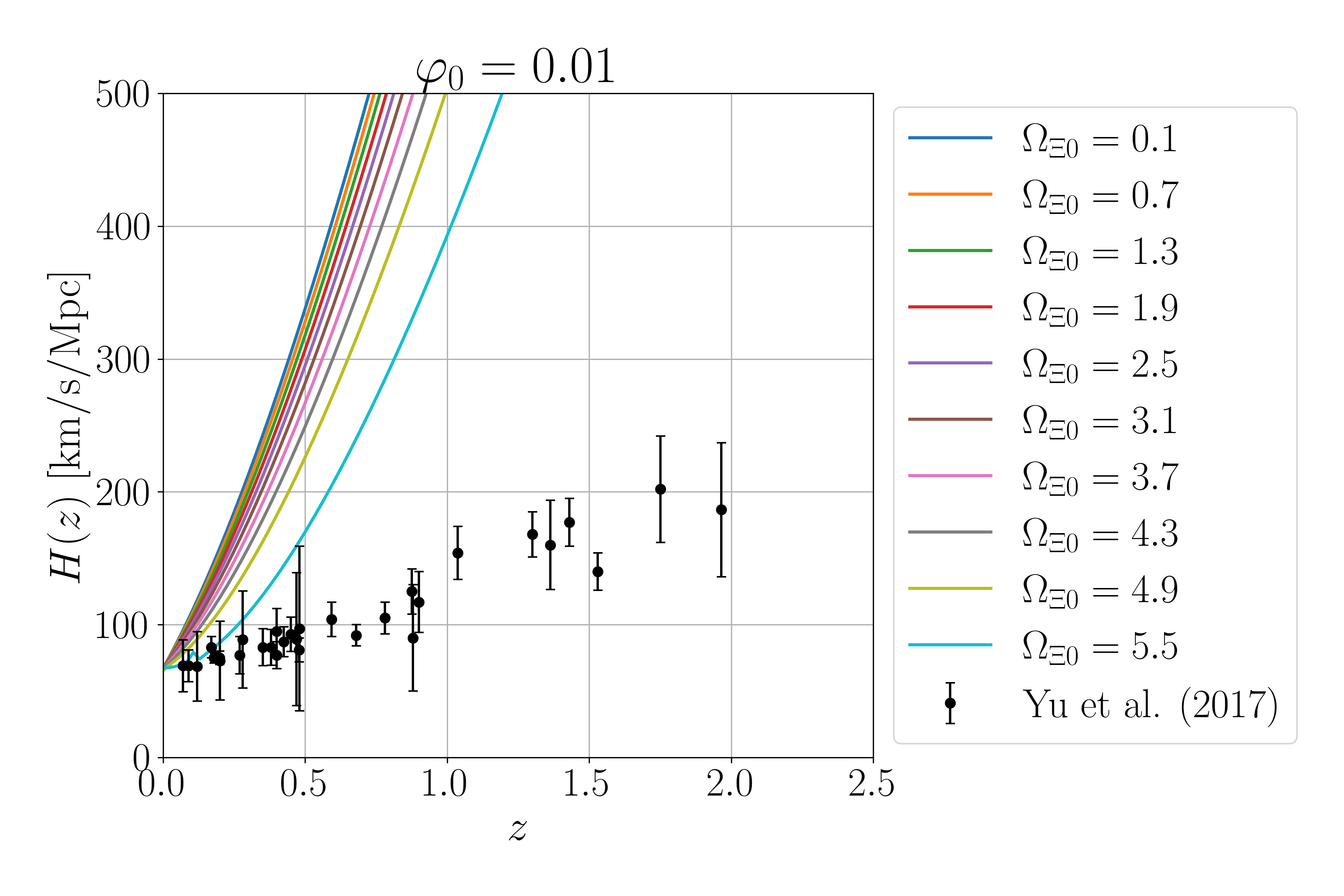}
        \includegraphics[width=0.5\textwidth]{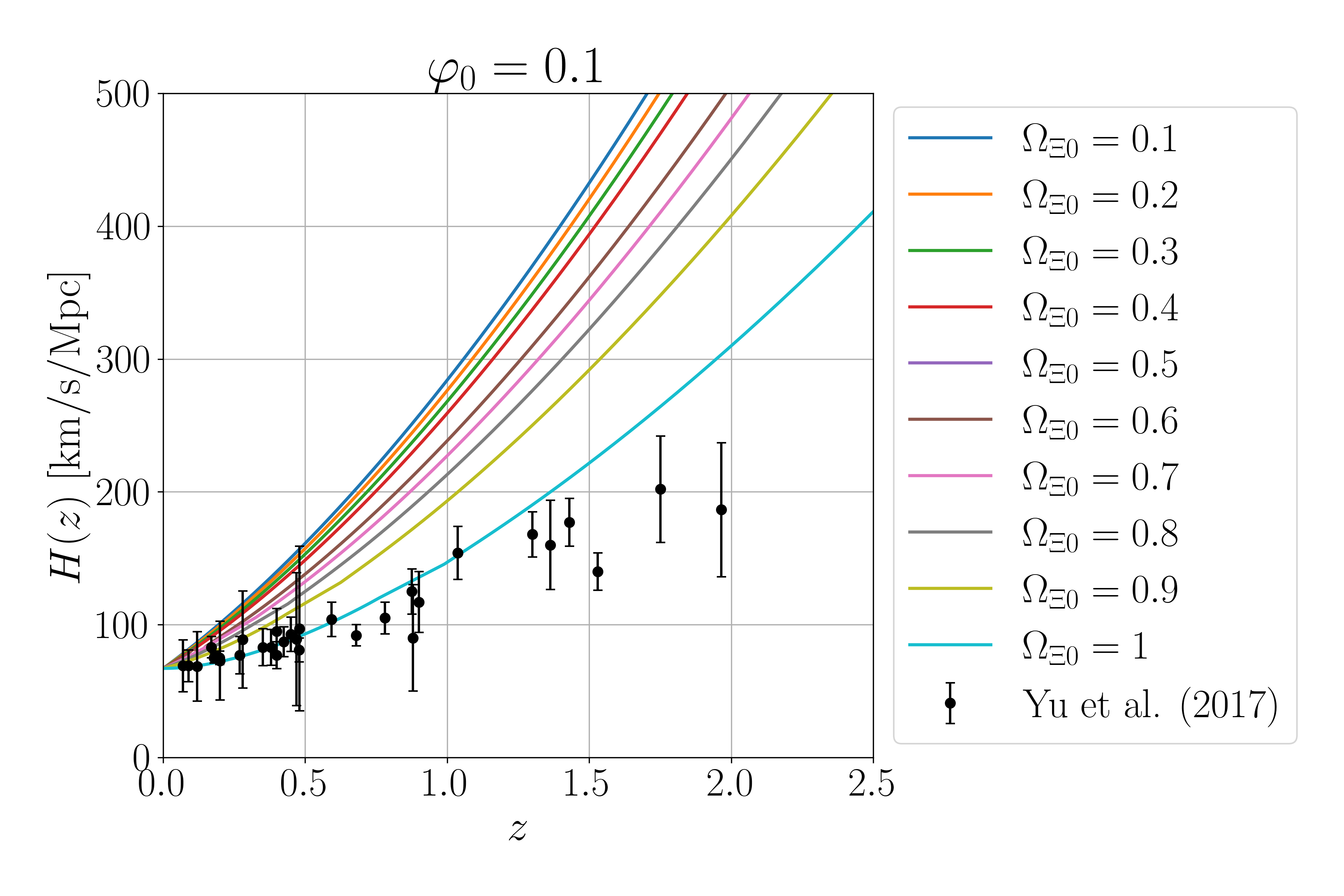} \\
        \includegraphics[width=0.5\textwidth]{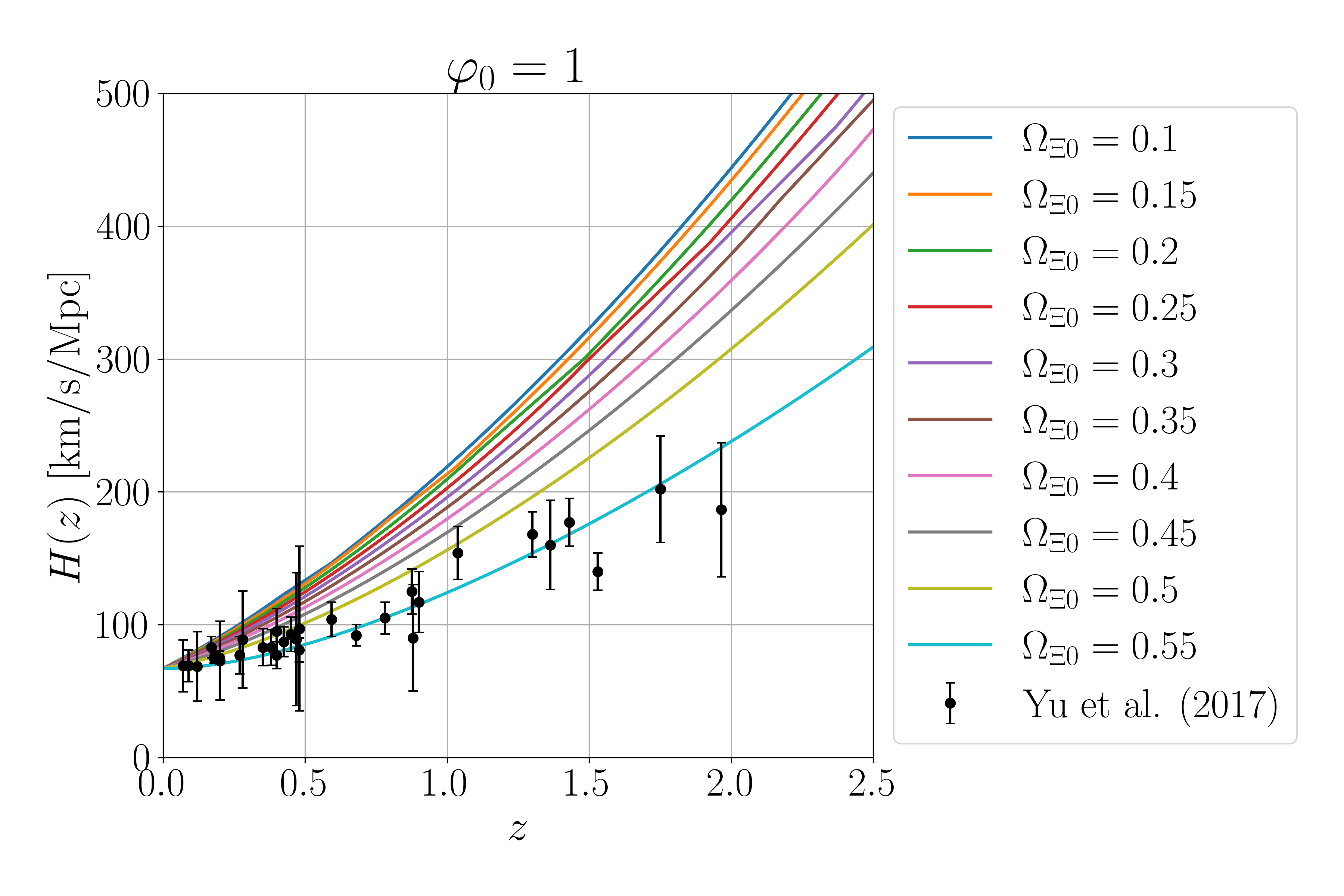}
        \includegraphics[width=0.5\textwidth]{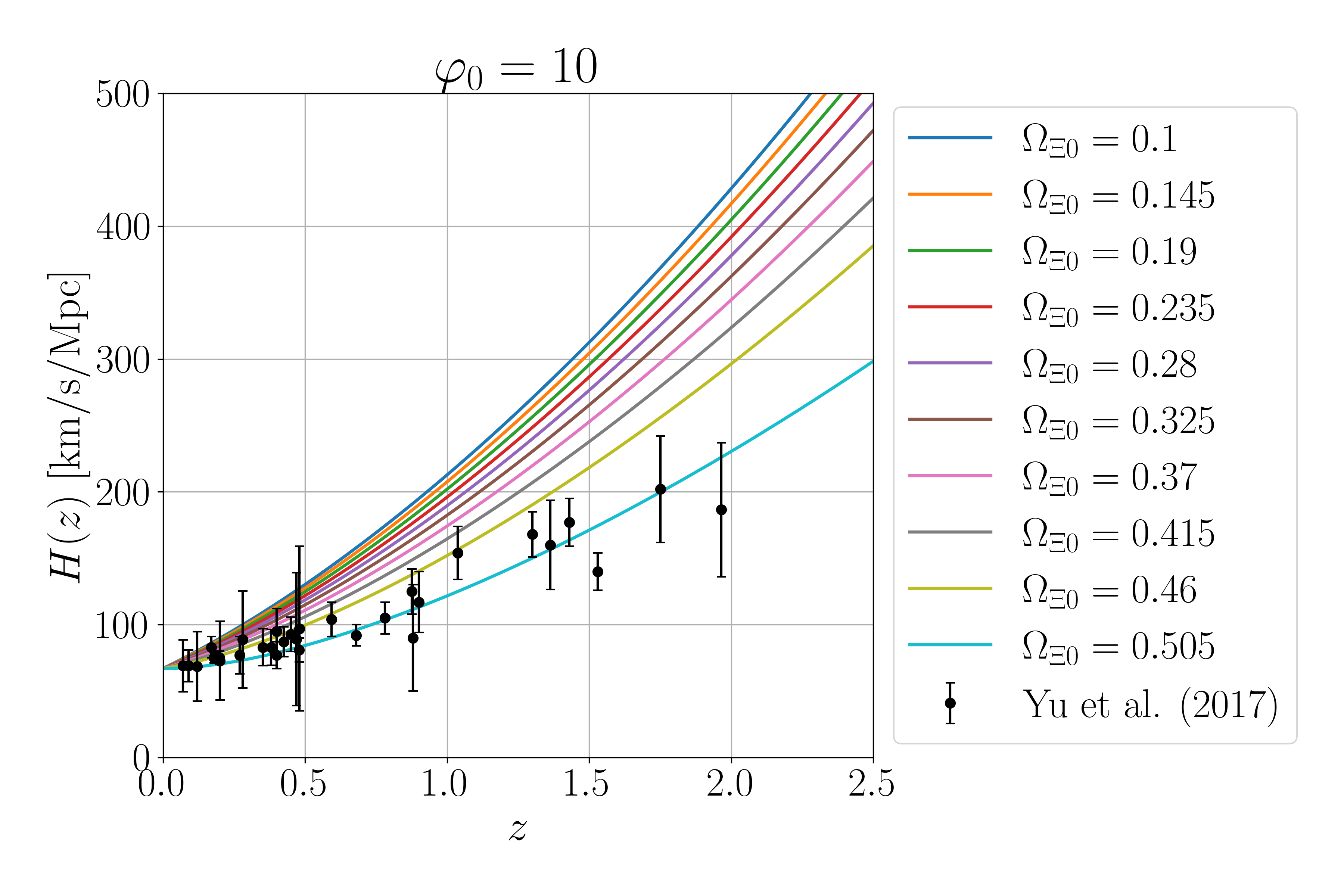} \\
        
    \caption{Redshift evolution of the Hubble-Lema\^\i tre parameter  $H(z)$ for different values of $(\varphi_0, \, \Omega_{\Xi 0})$ (solid curves). The solid points with error bars show the observational estimates from~\cite{Yu:2017iju}.}
    \label{effects of varying parameters}
\end{figure}

Figure \ref{effects of varying parameters} shows the effect of varying the model parameters on the $H(z)$ predictions. We considered different values of $\varphi_0$, or equivalently of $\Omega_0$, and let $\Omega_{\Xi 0}$ vary according to the condition of Eq. (\ref{omega_xi condition}). A visual inspection suggests that the curve that best describes the data, for each value of $\varphi_0$, seems to be systematically associated with the value of $\Omega_{\Xi 0}$ that saturates the upper bound in Eq.~(\ref{omega_xi condition}). This result is consistent with the one reported in~\cite{Sanna}, which was obtained by fitting the distance moduli of observed type Ia supernovae with the CRG model. Table \ref{chi sqrs} shows the $\chi^2$ values obtained when comparing the data points with the numerical curve for each value of $\varphi_0$ and the corresponding maximal value of $\Omega_{\Xi 0}$. 

Figure \ref{effects of varying parameters} and Table \ref{chi sqrs} show that values of $\varphi_0$ larger than 1 are favored over values lower than 1. This result is somewhat in contrast to expectations, since the gravitational boost mechanism described in earlier sections requires a small value of the gravitational permittivity $\epsilon(\rho)$, or equivalently of $\varphi$, to explain the missing mass problem on small scales. If the background value of $\varphi$ is larger than 1, and $\varphi$ is a monotonically increasing function of the local density, $\varphi$ will be larger than 1 on the scale of galaxies, where the local density is larger than the background cosmic density: these large values of $\varphi$ would be unable to solve the mass discrepancy problem. 

\begin{table}[!h]
\centering
\begin{tabular}{ll}
\toprule
($\varphi_0, \Omega_{\Xi 0}$) & $\chi^2$ \\
\midrule
(0.01, 5.5) & 4910 \\
(0.1, 1) & 71.22 \\
(1, 0.55) & 22.44  \\
(10, 0.505) & 21.87 \\
\bottomrule
\end{tabular}
\caption{$\chi^2$ values for selected pairs of the free parameters $\varphi_0$ and $\Omega_{\Xi 0}$. Each pair corresponds to the closest curve to the data points in each panel of Figure \ref{effects of varying parameters}. The underlying $\chi^2$ distribution has 29 degrees of freedom.}
\label{chi sqrs}
\end{table}

To further investigate this problem, we conducted a more thorough analysis based on the Monte Carlo Markov Chain (MCMC) method. We started with flat priors on the allowed parameter space ($0 < \varphi_0 < \varphi_{0,\text{max}}$, $0 < \Omega_{\Xi 0} \leq \left(\Omega_0+1\right)/2$) and evolved the chain with Gaussian steps of adaptive step size. The value of $\varphi_{0,\text{max}}$ has been varied during the analysis to properly study the parameter space, both above and below $\varphi_0 = 1$. We adopted the log-likelihood function
\begin{equation}
    \log \mathcal{L}\left(\varphi_0, \Omega_{\Xi 0}\right) = -\frac{1}{2}\sum_{i=1}^{N}\frac{\left[ H_{\text{mod}}(z_i;\varphi_0, \Omega_{\Xi 0}) - H_{\text{data}}(z_i)\right]^2}{\sigma^2_{H, \text{data}}}\,,
\end{equation}
to evaluate the goodness of the proposed step. If, at a certain step $n$ of the chain, with likelihood $\log \mathcal{L}_n = \log \mathcal{L}\left[(\varphi_0, \Omega_{\Xi 0})_n\right]$, the proposed step $(\varphi_0, \Omega_{\Xi 0})_{\text{prop}}$ satisfies $\log\mathcal{L}_{\text{prop}} > \log\mathcal{L}_n$, then we assign $(\varphi_0, \Omega_{\Xi 0})_{n+1} = (\varphi_0, \Omega_{\Xi 0})_{\text{prop}} $. Otherwise, we assign $(\varphi_0, \Omega_{\Xi 0})_{n+1} = (\varphi_0, \Omega_{\Xi 0})_{\text{prop}} $ with probability $P = \exp \left( \log \mathcal{L}_{\text{prop}} - \log\mathcal{L}_n \right)$ and $(\varphi_0, \Omega_{\Xi 0})_{n+1} = (\varphi_0, \Omega_{\Xi 0})_n $ with probability $1-P$. 
Finally, we verify the convergence using the Gelman-Rubin diagnostic $\cal{R}$~\cite{Gelman:1992zz}.
\begin{figure}[!htbp]
\begin{center}
        \includegraphics[width=0.49\textwidth]{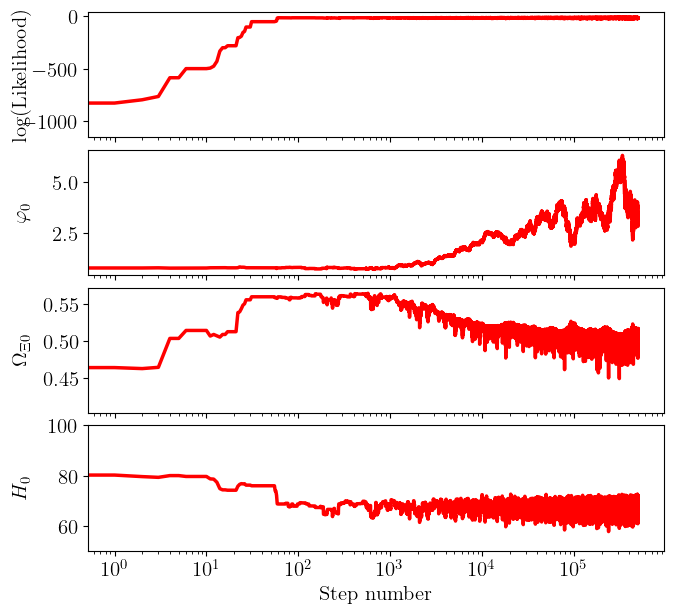}
        \includegraphics[width=0.49\textwidth]{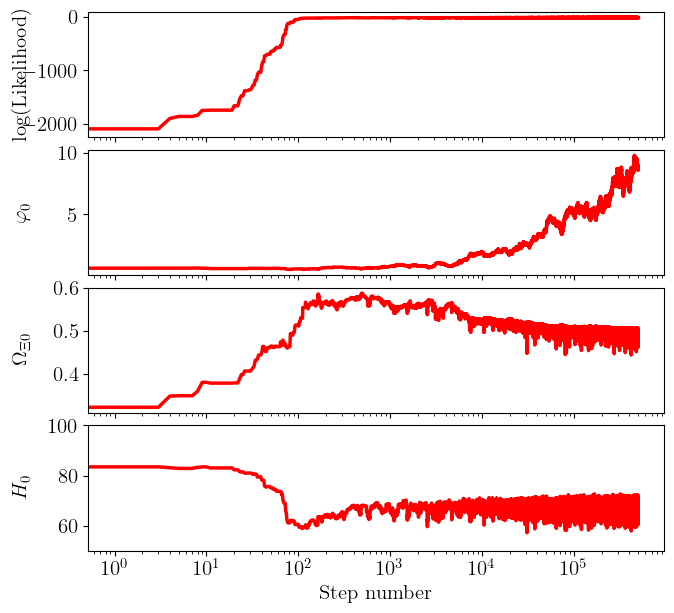} \\
        \includegraphics[width=0.49\textwidth]{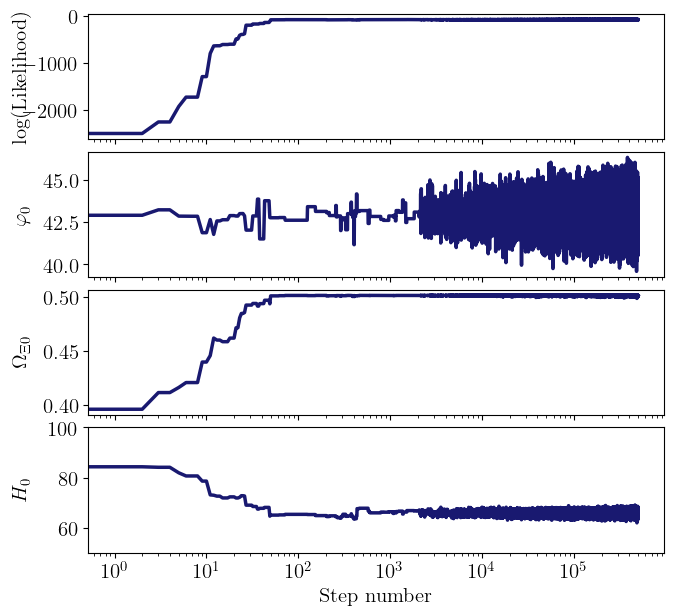}
        \includegraphics[width=0.49\textwidth]{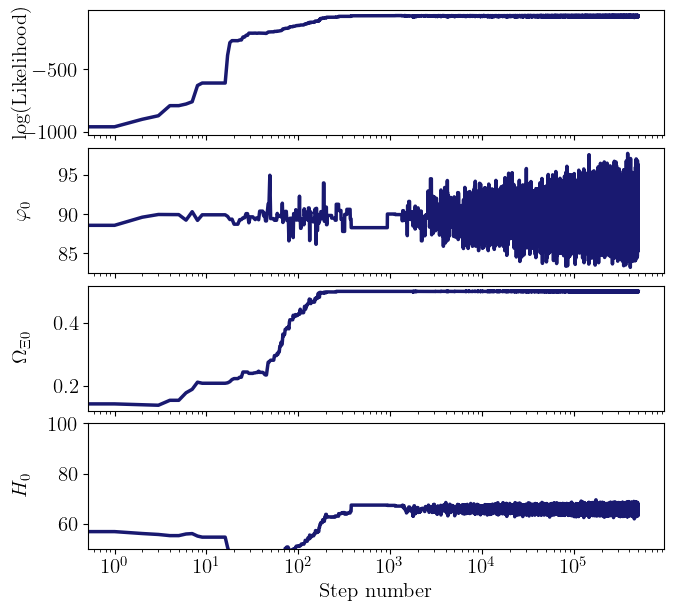} \\
        \includegraphics[width=0.49\textwidth]{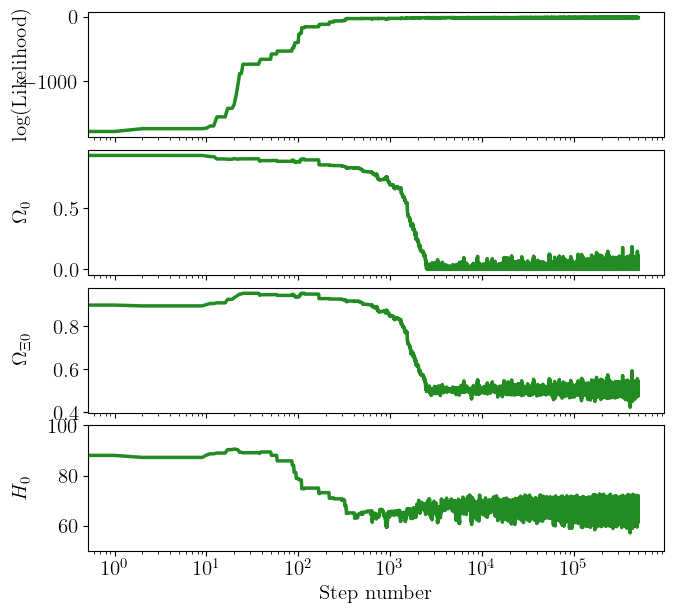}
        \includegraphics[width=0.49\textwidth]{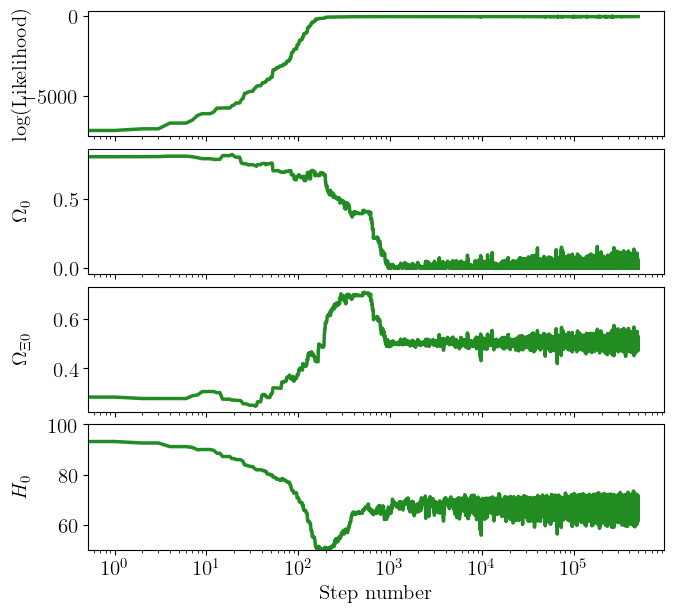} \\
    \end{center}
    \caption{Examples of the MCMC chains. We show six sets of four panels each.  Each set depicts the evolution of a single chain, including the burn-in set. The upper and middle sets of panels show the pair of free parameters $(\varphi_0, \Omega_{\Xi 0})$, with the Hubble-Lema\^\i tre parameter $H_0$  added for a consistency check. The prior for the value of $\varphi_0$ is set uniformly in the range $[0, 100]$. In the panels with the red curves, the initial value for $\varphi_0$ is chosen uniformly in the range $0<\varphi_0<1$, whereas in the middle sets (panels with the blue curves) this range was $1<\varphi_0<100$. The two lower sets with the green curves show the pair $(\Omega_0, \Omega_{\Xi 0})$.}
    \label{MCMC examples}
\end{figure}

Figure \ref{MCMC examples} shows some examples of the MCMC chains, including the burn-in sets to show the full evolution of the parameters along the chains. We fix the value of $\varphi_{0, \text{max}}$ above 1. However, if we select the initial value of $\varphi_0$ below unity (upper panels with red curves), the chains never converge in the space $\varphi_{0}<1$. On the contrary, if  we set the initial value of $\varphi_0$ larger than 1 (middle panels with blue curves), there appears to be no preferential value for $\varphi_0$, but rather each chain converges around its initial random value. This behavior of the chain suggests that the data cannot constrain the value of $\varphi_0$, provided that it is large enough.
Consequently, the value of $\Omega_0$, which is proportional to $\varphi_0^{-1}$, is also poorly constrained, although the value of $\Omega_0$ is in any case pushed towards zero because the chains move around values of $\varphi_0$ much larger than 1 (lower panels with green curves). For the Gelman-Rubin diagnostic $\cal{R}$, which tends to 1 when the chain converges~\cite{Gelman:1992zz}, we generally find ${\cal R}-1<10^{-3}$ when $\Omega_0$ reaches a value close zero.

\begin{figure}[!t]
    \begin{center}
         \includegraphics[width=0.75\textwidth]{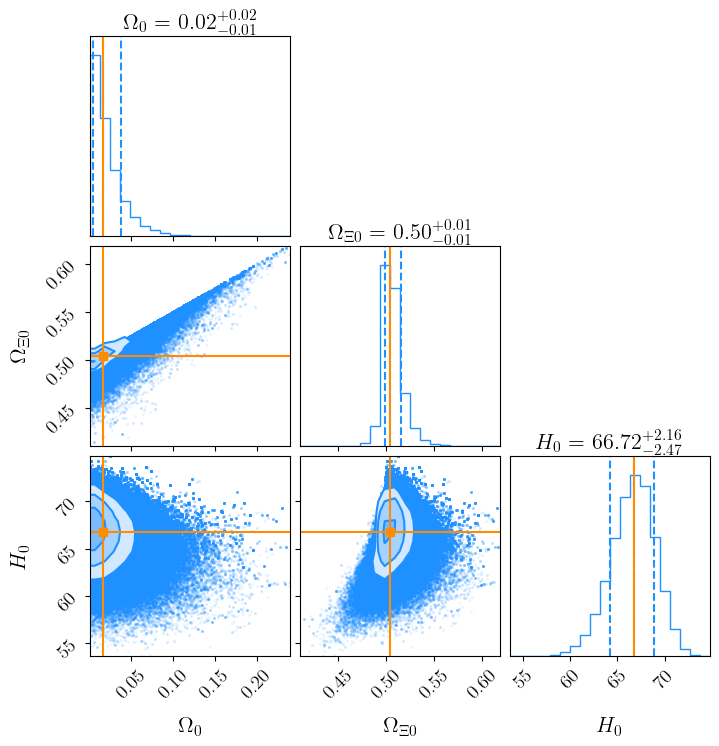}
    \end{center}

    \caption{Posterior distributions of the free parameters $\Omega_0$, $\Omega_{\Xi 0}$ and $H_0$ from the MCMC analysis. The median and the 0.16 and 0.84 quantiles are reported for each distribution (orange lines and dashed blue lines, respectively). In the $(\Omega_0, \Omega_{\Xi 0})$ parameter space, the chains  preferentially populate the region adjacent to the upper limit of $\Omega_{\Xi 0}$, with the preferred value for $\Omega_0$ consistent with zero at $2\sigma$.}
    \label{MCMC corner}
\end{figure}

In fact, Figure \ref{MCMC corner} shows the posterior distributions for the free parameters $\Omega_0$, $\Omega_{\Xi 0}$ and $H_0$. In this figure, we exclude an initial burn-in period, estimated for each chain via its autocorrelation time. We see once again how the most likely region of the parameter space is the one that saturates the bound in Eqs.~(\ref{omega_xi condition}): in fact, this plot very much resembles Figure (E.1) of~\cite{Sanna}. It is also clear that the preferred value of $\Omega_0$ is close to zero and, correspondingly, $\Omega_{\Xi 0} \sim 0.5$.

\section{Linear perturbation theory in the Brans--Dicke universe}
\label{CRG linear perturbation theory}
To investigate the formation of cosmic structure in a universe governed by our Brans--Dicke theory, we apply the linear perturbation theory to derive the perturbed versions of the Einstein field equations, the KG equation, and the continuity and Euler equations.  
We separately consider scalar and tensor perturbations, since, at linear order, these kinds of perturbation do not mix. The scalar perturbations describe the formation of the large-scale cosmic structure, whereas the tensor perturbations are potentially useful to study the propagation of gravitational waves. Below, we derive the equation for the evolution of the growth factor in a Brans--Dicke universe, and we solve it numerically. 

Our Brans--Dicke theory replaces the dark matter and dark energy of the standard model with a single scalar field. However, in our model, the scalar field is not expected to cluster as it happens in general UDM models~\cite{Bertacca:2008uf}. In these latter models, the scalar field acts as an effective dark fluid, that, by clustering, provides the gravitational potential wells where the baryonic matter falls. Indeed, a conformal transformation from the Jordan to the Einstein frame of our Brans--Dicke theory shows that our theory is equivalent to GR with an additional quintessence field, for which the perturbation sound speed is luminal. 
Therefore, the scalar field perturbations propagate very quickly and thus the scalar field clustering is suppressed. As a consequence, in our Brans--Dicke universe, the clustering of the baryonic matter is expected to be driven by the boost of the gravity force generated by the small values of the scalar field rather than by the clustering of the scalar field [see Eq.~(\ref{RG equation})].

In the following, we derive the sound speed of the scalar field, the evolution equations of the scalar perturbations, and the growth factor equation. Finally, we compute  the Laplacians of the two gravitational potentials of the perturbed FLRW metric.

\subsection{Sound speed of the scalar field}
\label{sec:soundspeed}

Here we confirm that the sound speed of the scalar field $c_{s,\delta\varphi}$  coincides with the speed of light, as anticipated above. Therefore, we do not expect the scalar field to cluster. In addition, this calculation is useful to check the sign of the square of the sound speed: a positive value of $c_{s,\delta\varphi}^2$ prevents the formation of gradient instabilities in the growth factor equation~\cite{Zumalacarregui:2016pph}, which would lead to an exponentially growing density contrast, thus breaking the validity of the assumptions of the linear perturbation theory. 

The expression for $c_{s,\delta\varphi}^2$ is known for a general Horndeski theory~\cite{Pace2019, Gleyzes2014}, and it can be computed from the $G_i$ functions of the given theory, which, in our case, are listed in Eq.~(\ref{Horndeski functions of CRG}). It is useful to adopt the following perturbation coefficients (we refer the reader to Eqs.~(B.1)-(B.5) of~\cite{Pace2019} for the complete definitions):
\begin{equation}
	\begin{aligned}
		M^2 & = 2G_4 = M_{\text{pl}}^2\varphi\,, \\
		\alpha_M & = \frac{1}{H}\frac{\mathrm{d} \ln M^2}{\mathrm{d} t} = \frac{\dot{\varphi}}{H\varphi}\,, \\
		\alpha_K & = \frac{1}{M^2H^2}2XG_{2,X} = - \frac{W(\varphi)}{H^2\varphi^2}X = \frac{X}{H^2\varphi^2} = - \frac{\dot{\varphi}^2}{H^2\varphi^2}\,, \\
		\alpha_B & = \frac{1}{2M^2H^2}2\dot{G}_4 = \frac{\dot{\varphi}}{2H\varphi}\,, \\
		\alpha_H & = \alpha_T = 0\,.
		\label{alphas}
	\end{aligned}
\end{equation}
With these coefficients, the general expression of the sound speed of the scalar field is~\cite{Pace2019, Gleyzes2014}
\begin{equation}
	c_{s,\delta\varphi}^2 = - \frac{2\left(1+\alpha_B\right)\left[\alpha_B\left(1+\alpha_T\right) - \left(\alpha_M - \alpha_T\right) + \frac{\dot{H}}{H^2}\right] + 2\dot{\alpha_B}H^{-1}}{\alpha} - \frac{\rho_{\rm m}\left(1+ w_{\rm m}\right)}{\alpha M^2H^2}\,,
	\label{cs2}
\end{equation}
where $\alpha = \alpha_K + 6\alpha_B^2$, and $\rho_{\mathrm{m}}$ and $w_{\mathrm{m}}$ refer to the matter component of the universe. By using in Eq.~(\ref{cs2}) the relations in Eqs.~(\ref{alphas}) and the analytical results of Eqs.~(\ref{phi analitico}) and~(\ref{dH/dt}), we find
\begin{equation}
	c_{s,\delta\varphi}^2 = \frac{6\dot{H} + 12H^2 + 6\Omega_0H_0H - 6\Omega_{\Xi 0}H_0^2}{6\dot{H} + 12H^2 + 3\Omega_0H_0H - 6\Omega_{\Xi 0}H_0^2} - \frac{3\Omega_0H_0H}{6\dot{H} + 12H^2 + 3\Omega_0H_0H - 6\Omega_{\Xi 0}H_0^2} = 1\, , 
	\label{cs2 final}
\end{equation}
as expected.

\subsection{Scalar perturbations}
We start by considering the scalar perturbations to a spatially-flat  FLRW background in the Newtonian gauge, i.e.
\begin{equation}
    \mathrm{d} s^{2}=a^{2}(\eta)\left[-(1+2 \Psi) \mathrm{d} \eta^{2}+(1+2 \Phi) \delta_{i j} \mathrm{d} x^{i} \mathrm{d} x^{j}\right]\,,
\end{equation}
where $\eta$ is the conformal time given by $ \mathrm{d} \eta = \mathrm{d}t/a(t)$, and $\Psi$ and $\Phi$ are the two independent gravitational potentials. Thus, to first order, the inverse metric $g^{\mu\nu} = \bar g^{\mu\nu} + \delta g^{\mu\nu} + \mathcal{O}(\delta g^2)$ has components  
\begin{equation}
\bar{g}^{00}=-a^{-2} ; \hspace{5mm} \bar{g}^{i j}=a^{-2} \delta^{i j} ; \hspace{5mm} \delta g^{00}=2 a^{-2} \Psi ; \hspace{5mm} \delta g^{i j}=-2 a^{-2} \Phi \delta^{i j}\,.
\end{equation}
We contract the modified Einstein equation [Eq.~(\ref{Rearrenged Modified Einsten})] with $g^{\mu\lambda}$ and perturb it to find
\begin{align*}
\delta & \varphi R_{\enspace\nu}^{\mu} + \varphi \delta R_{\enspace\nu}^{\mu} - \delta g^{\mu \lambda} \partial_{\lambda} \partial_{\nu} \varphi - g^{\mu \lambda} \partial_{\lambda} \partial_{\nu} \delta \varphi + \delta g^{\mu \lambda} \Gamma_{\enspace\lambda \nu}^{\rho} \partial_{\rho} \varphi + g^{\mu \lambda} \delta \Gamma_{\enspace\lambda \nu}^{\rho} \partial_{\rho} \varphi + g^{\mu \lambda} \Gamma_{\enspace\lambda \nu}^{\rho} \partial_{\rho} \delta \varphi + \\
& - \frac{\delta \varphi}{\varphi^{2}} g^{\mu \lambda} \partial_{\lambda} \varphi \partial_{\nu} \varphi + \frac{1}{\varphi} \delta g^{\mu \lambda} \partial_{\lambda} \varphi \partial_{\nu} \varphi + \frac{1}{\varphi} g^{\mu \lambda} \partial_{\lambda} \delta \varphi \partial_{\nu} \varphi + \frac{1}{\varphi} g^{\mu \lambda} \partial_{\lambda} \varphi \partial_{\nu} \delta \varphi = 8 \pi G \delta T_{\enspace\nu}^{\mu}\,, \numberthis
\label{perturbed einstein}
\end{align*}
where the perturbed Christoffel symbols to first order are 
\begin{equation}
    \delta \Gamma_{\enspace \mu \nu}^{\lambda}=\frac{1}{2} \delta g^{\lambda \alpha}\left(\partial_{\mu} g_{\nu \alpha}+\partial_{\nu} g_{\mu \alpha}-\partial_{\alpha} g_{\mu \nu}\right)+\frac{1}{2} g^{\lambda \alpha}\left(\partial_{\mu} \delta g_{\nu \alpha}+\partial_{\nu} \delta g_{\mu \alpha}-\partial_{\alpha} \delta g_{\mu \nu}\right)\,,
\end{equation}
while the perturbed Ricci tensor components are
\begin{equation}
    \delta R_{\mu \nu}=\partial_{\nu} \delta \Gamma_{\enspace \mu \lambda}^{\lambda}-\partial_{\lambda} \delta \Gamma_{\enspace \mu \nu}^{\lambda}+\delta \Gamma_{\enspace \mu \rho}^{\lambda} \Gamma_{\enspace \lambda \nu}^{\rho}+\Gamma_{\enspace \mu \rho}^{\lambda} \delta \Gamma_{\enspace \lambda \nu}^{\rho}-\delta \Gamma_{\enspace \mu \nu}^{\lambda} \Gamma_{\enspace\lambda \rho}^{\rho}-\Gamma_{\enspace \mu \nu}^{\lambda} \delta \Gamma_{\enspace \lambda \rho}^{\rho}\,,
\end{equation}
and consequently $\delta R_{\enspace\nu}^{\mu}=\delta g^{\mu \alpha} R_{\alpha \nu}+g^{\mu \alpha} \delta R_{\alpha \nu}$. Finally, the perturbed components of the stress-energy tensor are 
\begin{equation}
    \delta T_{\mu \nu}=(\delta \rho+\delta P) u_{\mu} u_{\nu}+(\rho+P)\left[\delta u_{\mu} u_{\nu}+u_{\mu} \delta u_{\nu}\right]+\delta P g_{\mu \nu}+P \delta g_{\mu \nu}+\pi_{\mu \nu}\,,
\end{equation}
where $\pi_{\mu\nu}$ is the anisotropic stress and $u^\mu$ is the 4-velocity of fluid particles, whose normalization $u^\mu u_\mu = -1 $ yields
\begin{equation}
    u^{\mu}=\left[\frac{1}{a}(1-\Psi), \frac{v^{i}}{a}\right]\,,
\end{equation}
where $v^i$ is the peculiar velocity of the particle. Again, $\delta T_{\enspace \nu}^{\mu}=\delta g^{\mu \alpha} T_{\alpha \nu}+g^{\mu \alpha} \delta T_{\alpha \nu}$. 

By using all these results, denoting with the prime the derivation with respect to the conformal time, we find, for the $(00)$ component of Eq.~(\ref{perturbed einstein}),
\begin{align*}
    \varphi & \left[6 \mathcal{H}^\prime \Psi-3 \Phi^{\prime\prime}+\nabla^{2} \Psi-3 \mathcal{H}(\Phi^{\prime}-\Psi^\prime)\right]-3 \mathcal{H}^\prime \delta \varphi+2 \Psi \varphi^{\prime\prime}-\delta \varphi^{\prime\prime}-2 \mathcal{H} \varphi^\prime \Psi+ \varphi^\prime \Psi^\prime + \mathcal{H} \delta \varphi^\prime\\
    &-\frac{\delta \varphi}{\varphi^{2}} \left(\varphi^\prime\right)^{2}-\frac{2 \left(\varphi^\prime\right)^{2}}{\varphi} \Psi +\frac{2 \varphi^\prime}{\varphi} \delta \varphi^\prime=8 \pi G a^{2} \delta \rho\,, \numberthis
    \label{00 perturbed einstein}
\end{align*}
where $\mathcal{H}=a^\prime/a$. For the $(0i)$ components, we find
\begin{equation}
    2\varphi\left(\partial_i\Phi^\prime - \mathcal{H}\partial_i\Psi\right) + \partial_i\delta\varphi^\prime - \partial_i\Psi\varphi^\prime - \mathcal{H}\partial_i\delta\varphi - \frac{\varphi^\prime}{\varphi}\partial_i\delta\varphi = 8\pi G a^2\left(\rho + P\right)v_i\,,
    \label{0i perturbed einstein}
\end{equation}
and, finally, for the $(ij)$ components, we find
\begin{align*}
    & \left\{-\delta \varphi\left(\mathcal{H}^\prime+2 \mathcal{H}^{2}\right)+\varphi\left(\nabla^{2} \Phi+2 \mathcal{H}^\prime \Psi-\Phi^{\prime\prime}+\mathcal{H} \Psi^\prime-5 \mathcal{H} \Phi^\prime + 4 \mathcal{H}^{2} \Psi\right)+2 \Phi \mathcal{H} \varphi^\prime \right. \\
    & \quad \left. -\varphi^\prime\left[2 \mathcal{H}(\Phi-\Psi)+ \Phi^\prime\right] -\mathcal{H} \delta \varphi^\prime\right\} \delta_{\enspace j}^{i} + \varphi \partial^{i} \partial_{j}\left(\Psi+\Phi+\frac{\delta \varphi}{\varphi}\right)=-8 \pi G a^{2}\left(\delta P \delta_{\enspace j}^{i}+\pi_{j}^{i}\right)\,. \numberthis
    \label{ij perturbed einstein}
\end{align*}
The latter can be split into an equation for the traceless part
\begin{equation}
    \partial^{i} \partial_{j}\left(\Psi+\Phi+\frac{\delta \varphi}{\varphi}\right)=-\frac{8 \pi G}{\varphi} a^{2} \partial^{i} \partial_{j} \Pi \,;
\end{equation}
and an equation for the trace part 
\begin{align*}
    & \left\{-\delta \varphi\left(\mathcal{H}^\prime+2 \mathcal{H}^{2}\right)+\varphi\left(\nabla^{2} \Phi+2 \mathcal{H}^\prime \Psi-\Phi^{\prime\prime}+\mathcal{H} \Psi^\prime-5 \mathcal{H} \Phi^\prime + 4 \mathcal{H}^{2} \Psi\right)+2 \Phi \mathcal{H} \varphi^\prime \right.\\
    & \left. \quad - \varphi^\prime[2 \mathcal{H}(\Phi-\Psi)+ \Phi^\prime]\right. -\mathcal{H} \delta \varphi^\prime\}+\frac{\varphi}{3} \nabla^{2}\left(\Psi+\Phi+\frac{\delta \varphi}{\varphi}\right)=-8 \pi G a^{2} \delta P \,, \numberthis
\end{align*}
where $\Pi$ is the scalar function that characterizes the spatial part of the anisotropic stress tensor, i.e., $\pi_{ij} = a^2\left(\partial_i\partial_j\Pi - \delta_{ij}\nabla^2\Pi/3\right)$.

Next, we perturb the KG equation, Eq.~(\ref{Rearrenged KG}), which gives
\begin{align*}
    & \delta g^{\alpha \beta} \partial_{\alpha} \partial_{\beta} \varphi+g^{\alpha \beta} \partial_{\alpha} \partial_{\beta} \delta \varphi-\delta g^{\alpha \beta} \Gamma_{\alpha \beta}^{\lambda} \partial_{\lambda} \varphi-g^{\alpha \beta} \delta \Gamma_{\alpha \beta}^{\lambda} \partial_{\lambda} \varphi-g^{\alpha \beta} \Gamma_{\alpha \beta}^{\lambda} \partial_{\lambda} \delta \varphi \\
    & \quad -2\Xi\delta\varphi = 8\pi G \left( \delta g^{\mu \nu} T_{\mu \nu}+g^{\mu \nu} \delta T_{\mu \nu}\right)\,, \numberthis
\end{align*}
which becomes, using all the previous results,
\begin{equation}
    2 \Psi \varphi^{\prime\prime}-\delta \varphi^{\prime}+\nabla^{2} \delta \varphi+4 \mathcal{H} \varphi^\prime \Psi+ \varphi^\prime ( \Psi^\prime-3 \Phi^\prime)-2 \mathcal{H} \delta^\prime \varphi-2 \Xi a^{2} \delta \varphi=8 \pi G a^{2}(3 \delta P-\delta \rho) \,.
    \label{perturbed KG}
\end{equation}
Finally, we also perturb the continuity equation, $\nabla_\mu T^\mu_{\enspace \nu} = 0$, obtaining
\begin{equation}
    \partial_{\mu} \delta T_{\enspace \nu}^{\mu}-\delta \Gamma_{\enspace\mu \nu}^{\lambda} T_{\enspace\lambda}^{\mu}-\Gamma_{\enspace \mu \nu}^{\lambda} \delta T_{\enspace\lambda}^{\mu}+\delta \Gamma_{\enspace\mu \lambda}^{\mu} T_{\enspace\nu}^{\lambda}+\Gamma_{\enspace\mu \lambda}^{\mu} \delta T_{\enspace\nu}^{\lambda}=0\,,
\end{equation}
which gives, for $\nu = 0$,
\begin{equation}
    \delta \rho^\prime + 3 \mathcal{H}(\delta \rho+\delta P)+(\rho+P)\left(\partial_{i} v^{i} + 3\Phi^\prime\right)=0 \,;
\end{equation}
and, for $\nu = i$,
\begin{equation}
    (\rho+P)\left(4 \mathcal{H} v_{i}+v^\prime_{i}+\partial_{i} \psi\right)+v_{i}(\rho^\prime+ P^\prime)+\partial_{i} \delta P+\partial_{j} \pi_{i}^{j}=0 \,.
\end{equation}

It is convenient to recast some of these results using the equation of state $P = w\rho$ and the definitions for the speed of sound $c_s^2 = \delta P/\delta\rho$, the density contrast $\delta = \delta\rho/\rho$ and the velocity divergence $\theta = \partial_i v^i$. Then we obtain the following set of equations of motion:
\begin{subequations} \label{perturbed system}
\begin{equation}
\begin{aligned}
    \varphi & \left[6 \mathcal{H}^\prime \Psi-3 \Phi^{\prime\prime}+\nabla^{2} \Psi-3 \mathcal{H}(\Phi^{\prime}-\Psi^\prime)\right]-3 \mathcal{H}^\prime \delta \varphi+2 \Psi \varphi^{\prime\prime}-\delta \varphi^{\prime\prime}-2 \mathcal{H} \varphi^\prime \Psi\\
    &+ \varphi^\prime \Psi^\prime + \mathcal{H} \delta \varphi^\prime-\frac{\delta \varphi}{\varphi^{2}} \left(\varphi^\prime\right)^{2}-\frac{2 \left(\varphi^\prime\right)^{2}}{\varphi} \Psi +\frac{2 \varphi^\prime}{\varphi} \delta \varphi^\prime=8 \pi G a^{2} \rho\delta \,,
\end{aligned}
\label{perturbed system 00}
\end{equation}

\begin{equation}
   \nabla^2 \left[ 2\varphi\left(\Phi^\prime - \mathcal{H}\Psi\right) + \delta\varphi^\prime - \Psi\varphi^\prime - \mathcal{H}\delta\varphi - \frac{\varphi^\prime}{\varphi}\delta\varphi \right]= 8\pi G a^2\rho\left(1 + w\right)\theta\,,
\label{perturbed system 0i}
\end{equation}

\begin{equation}
    \Psi+\Phi+\frac{\delta \varphi}{\varphi}=-\frac{8 \pi G}{\varphi} a^{2} \Pi \,,
    \label{perturbed system ij traceless}
\end{equation}

\begin{equation}
\begin{aligned}
     & \left\{-\delta \varphi\left(\mathcal{H}^\prime+2 \mathcal{H}^{2}\right)+\varphi\left(\nabla^{2} \Phi+2 \mathcal{H}^\prime \Psi-\Phi^{\prime\prime}+\mathcal{H} \Psi^\prime-5 \mathcal{H} \Phi^\prime + 4 \mathcal{H}^{2} \Psi\right)+2 \Phi \mathcal{H} \varphi^\prime \right.\\
    & \left. \quad - \varphi^\prime[2 \mathcal{H}(\Phi-\Psi)+ \Phi^\prime]\right. -\mathcal{H} \delta \varphi^\prime\}+\frac{\varphi}{3} \nabla^{2}\left(\Psi+\Phi+\frac{\delta \varphi}{\varphi}\right)=-8 \pi G a^{2} c_s^2\rho\delta\,,
\end{aligned}
\label{perturbed system ij traceful}
\end{equation}

\begin{equation}
    2 \Psi \varphi^{\prime\prime}-\delta \varphi^{\prime\prime}+\nabla^{2} \delta \varphi+4 \mathcal{H} \varphi^\prime \Psi+ \varphi^\prime ( \Psi^\prime-3 \Phi^\prime)-2 \mathcal{H} \delta \varphi^\prime-2 \Xi a^{2} \delta \varphi=8 \pi G a^{2}\rho\delta(3 c_s^2-1) \,,
    \label{perturbed system KG}
\end{equation}

\begin{equation}
    \delta^\prime + 3 \mathcal{H}(c_s^2 - w)\delta +(1+w)\left(\theta + 3\Phi^\prime\right)=0 \,,
    \label{perturbed system continuity}
\end{equation}

\begin{equation}
    \rho(1+w)\left( \mathcal{H} \theta + \theta^\prime + \nabla^2 \Psi -3w\mathcal{H}\theta\right) + c_s^2\rho\nabla^2\delta + \frac{2}{3}\nabla^4\Pi = 0\,,
    \label{perturbed system Euler}
\end{equation}
\end{subequations}
where $\nabla^4\Pi = \nabla^2\nabla^2\Pi$. 

For completeness, we have also derived the equations of motion for tensor perturbations, potentially useful to describe the propagation of gravitational waves. They are reported in Appendix \ref{TensorPerturbations}.

\subsection{Growth factor equation}
\label{growth factor equation}
We now apply our system of perturbed equations, Eqs.~(\ref{perturbed system}), to derive the growth factor equation. Since we want to study the formation of structure at late times, pressure waves due to radiation are negligible and thus we assume $c_s^2 = w = 0$ for our cosmic fluid. Moreover, we can take the anisotropic stress, represented by $\Pi$, to be negligible. With these assumptions, the continuity and Euler equations, Eqs.~(\ref{perturbed system continuity}) and (\ref{perturbed system Euler}), simplify to
\begin{align}
    \delta^\prime = & - \left(\theta + 3\Phi^\prime\right)\,, \label{continuità}\\ 
    \theta^\prime = & - \mathcal{H}\theta - \nabla^2\Psi\,.   \label{eulero}
\end{align}
By multiplying by $a$ and differentiating with respect to $\eta$, Eq.~(\ref{continuità}) yields, together with Eqs.~(\ref{eulero}),~(\ref{perturbed system 00}) and~(\ref{perturbed system KG}), 
\begin{equation}
    \begin{split}
        \left(a\delta^\prime\right)^\prime  =& - a\left(6\mathcal{H}^\prime\Psi + 3\mathcal{H}\Psi^\prime\right) + \frac{8\pi Ga^3}{\varphi}\left(2\delta\rho - 3\delta P\right) + 3a\mathcal{H}^\prime\frac{\delta\varphi}{\varphi} + a\frac{\nabla^2\delta\varphi}{\varphi} + 6a\mathcal{H}\frac{\varphi^\prime}{\varphi}\Psi\\
        & - 3a\Phi^\prime\frac{\varphi^\prime}{\varphi} - 3a\mathcal{H}\frac{\delta\varphi^\prime}{\varphi} - 2\Xi a^3\frac{\delta\varphi}{\varphi} + a\frac{\delta\varphi}{\varphi^3}\left(\varphi^\prime\right)^2 + 2a\left(\frac{\varphi^\prime}{\varphi}\right)^2\Psi - 2a\frac{\varphi^\prime}{\varphi^2}\delta\varphi^\prime\,.
    \end{split}
    \label{passaggio 2}
\end{equation}
Compared to its $\Lambda$CDM counterpart, this equation features additional terms that depend on the scalar field and its perturbations. Moreover, there are terms that depend on the gravitational potentials $\Phi$ and $\Psi$. However, these terms are subdominant on scales much smaller than the cosmological horizon and, therefore, can be neglected.
Despite these simplifications, a complete solution must be found numerically, by coupling Eq.~(\ref{passaggio 2}) with Eqs.~(\ref{perturbed KG}) and~(\ref{modified KG}). 
We transform Eq.~(\ref{passaggio 2}) by switching to the independent variable $N\equiv\ln a$. We also move to Fourier space, with the standard decomposition of the perturbative quantities in normal modes of the wave-vector $\bf{k}$. Thus we get
\begin{equation}
    \begin{split}
         & \delta_{\textbf{k}}^{\prime\prime} + \left(2 + \frac{H^\prime}{H}\right)\delta_{\textbf{k}}^\prime - \frac{16\pi G}{\varphi}\frac{\rho_{{\rm m}0}}{H^2a^3}\delta_{\textbf{k}} + \frac{k^2}{a^2H^2}\frac{\delta\varphi_{\textbf{k}}}{\varphi} + 2\frac{\Xi}{H^2}\frac{\delta\varphi_{\textbf{k}}}{\varphi} \\
         & \quad = \frac{\delta\varphi_{\textbf{k}}}{\varphi^3}\left(\varphi^\prime\right)^2 - 2\frac{\varphi^\prime}{\varphi^2}\delta\varphi_{\textbf{k}}^\prime - 3\frac{\delta\varphi_{\textbf{k}}^\prime}{\varphi} + 3\frac{\delta\varphi_{\textbf{k}}}{\varphi} + 3\frac{H^\prime}{H}\frac{\delta\varphi_{\textbf{k}}}{\varphi}\,,
     \label{delta evolution}
     \end{split}
\end{equation}
where the prime now denotes differentiation with respect to $N$. Adopting the same numerical approach used in Section~\ref{Numerical background solution and $H(z)$ fit}, we solve  the background KG equation~[Eq.~(\ref{modified KG})], the perturbed KG equation~[Eq.~(\ref{perturbed KG})], and Eq.~(\ref{delta evolution}). 

\begin{figure}[htbp]
        \includegraphics[width=\textwidth]{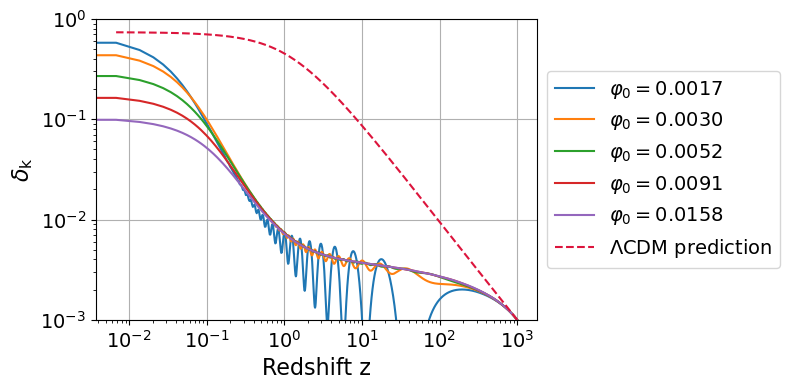}
        \includegraphics[width=\textwidth]{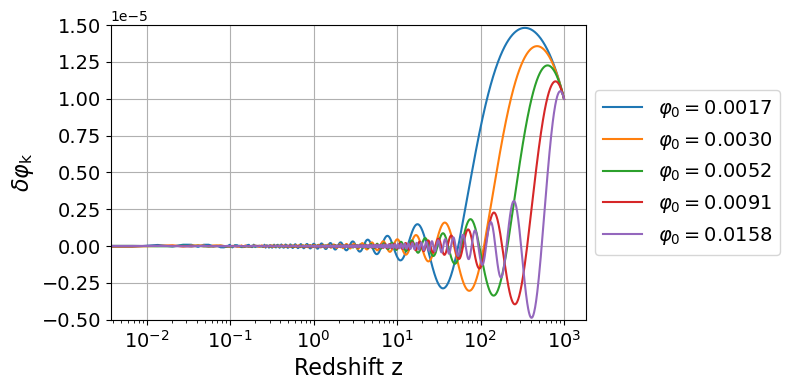}
        
    \caption{Evolution of the density contrast $\delta_{\textbf{k}}$ (top panel) and scalar field perturbations  $\delta\varphi_{\textbf{k}}$ (bottom panel) as a function of redshift, for $k = 1 \, \text{Mpc}^{-1}$. Different curves assume different values for the parameter $\varphi_0$; the other free parameter, $\Omega_{\Xi 0}$, was fixed to the upper limit of Eq.~(\ref{omega_xi condition}), that was shown to be the best-fit value for the background evolution, as explained in Section~\ref{Numerical background solution and $H(z)$ fit}. The dashed red line in the top panel represents the $\Lambda$CDM prediction.}
    \label{delta evolution figure}
\end{figure}
The upper panel of Figure \ref{delta evolution figure} shows the evolution of the density contrast $\delta_{\textbf{k}}$ as a function of redshift for a fixed wave-number, $k = 1 \, \text{Mpc}^{-1}$. Different curves assume slightly different values of the scalar field today, but the general range explored ($\varphi \lesssim 0.1$) has been chosen to obtain the largest growth possible. The results are highly dependent on the choice of the initial conditions for $\delta_{\textbf{k}}$ and $\delta\varphi_{\textbf{k}}$. Specifically, if the perturbations for the scalar field are taken to be of the same order as the density contrast, or even larger, the effects of the terms proportional to $\delta\varphi_{\textbf{k}}$ in Eq.~(\ref{delta evolution}) become sizable and the behavior of the density contrast can be modified to the point that the mass over-densities decrease with time. On the contrary, the mass over-densities increase with time if the initial $\delta\varphi_{\textbf{k}}$'s are set to a few orders of magnitude smaller than the initial $\delta_{\textbf{k}}$'s. In this case, the perturbations of the scalar field  are washed out because of their relativistic sound speed (lower panel of Figure \ref{delta evolution figure}) and all the terms proportional to $\delta\varphi_{\textbf{k}}$ and its derivative in Eq.~(\ref{delta evolution}) have thus very little effect on the behavior of the evolution of the mass overdensities. Figure \ref{delta evolution figure}  shows the time evolution of  $\delta_{\textbf{k}}$ and $\delta\varphi_{\textbf{k}}$  when the initial conditions are set to $\delta_{\textbf{k}}(z=1000) = 10^{-3}$ and $\delta\varphi_{\textbf{k}}(z=1000) = 10^{-5}$.

Figure \ref{delta evolution figure} shows that the model predicts a very delayed growth of the density contrast that reaches the non-linear regime at  redshifts $z<1$. This prediction is at odds with the observation of the most ancient galaxies, which are known to be present at redshifts as high as $z\sim 14$~\cite{naidu2025cosmicmiracleremarkablyluminous}. In contrast, the $\Lambda$CDM solution predicts a substantial growth of structure at larger redshifts, although the predicted number density of massive galaxies still appears to be an order of magnitude smaller than observed~\cite{Boylan-Kolchin:2011qkt, Labbe_2023}.  
This retarded evolution of the perturbations originates from the coefficient of the source term in Eq.~(\ref{delta evolution}), the third term on the left-hand side of the growth factor equation: $16\pi G\rho_{{\rm m}0}/(\varphi H^2a^3)$, with $\rho_{{\rm m}0}$ the density of the baryonic matter at the present time. According to Eq.~(\ref{phi analitico}), $\varphi\propto H^{-1}a^{-3}$, so the coefficient of the source term is proportional to $H^{-1}$ and decreases with increasing redshift. Figure \ref{comparison analytical numerical} shows that the Hubble parameter decreases with $a^{-\sqrt{3}}$ [Eq.~(\ref{H(a)})]. Therefore, the contribution of the source term is negligible at high redshift and the evolution of the perturbations can only start at relatively small redshift. We recall that in the $\Lambda$CDM model, this coefficient is basically constant, with $\rho_{{\rm m}0}$ the density of the baryonic and dark matter combined. We checked that even considering a constant $\Xi$ with opposite sign returns qualitative similar results and still predicts a retarded formation of the cosmic structure.

\subsection{Laplacians of the gravitational potentials}
\label{laplacian of the gravitational potentials}
An important application of structure formation is the study of gravitational lensing. Although we do not intend to provide a complete analysis of this topic in the context of our Brans--Dicke theory, we want to underline here an important aspect of it, namely how the choice of the scalar potentials $W(\varphi)$ and $V(\varphi)$ can impact the intensity of the gravitational lensing effect.

By taking a Brans--Dicke action with generic $W(\varphi)$ and $V(\varphi)$, the equations of motion for the metric are
\begin{align}
\begin{split}
    \varphi R_{\mu\nu} + g_{\mu\nu}\left[(W+1)\square\varphi + V - \varphi \frac{{\rm d}V}{{\rm d}\varphi} + \frac{1}{2}\frac{{\rm d}W}{{\rm d}\varphi}\nabla_\alpha\varphi\nabla^\alpha\varphi\right] & \\
     - \nabla_\mu\nabla_\nu\varphi - \frac{W}{\varphi} \nabla_\mu\varphi\nabla_\nu\varphi & = 8\pi GT_{\mu\nu}.
\end{split}
\label{EE generic BD}
\end{align}
This equation reduces to Eq.~(\ref{Rearrenged Modified Einsten}) for the choices of Eq.~(\ref{choices}). Perturbing Eq.~(\ref{EE generic BD}) and evaluating its $(00)$ component and the trace part of its $(ij)$ component gives, respectively,
    \begin{align}
        \begin{split}
            -\varphi\nabla^2\Phi - a^2\varphi\delta\varphi\frac{{\rm d}^2V}{{\rm d}\varphi^2} + (W+1)\nabla^2\delta\varphi & = 0\,;
        \end{split}\\
        \begin{split}
            -\varphi\nabla^2\Psi - a^2\varphi\delta\varphi \frac{{\rm d}^2V}{{\rm d}\varphi^2} + (W+1)\nabla^2\delta\varphi & = - 8\pi G\delta\rho\,,
        \end{split}
    \end{align}
where, once again, we have neglected terms proportional to the time derivatives of $\Phi$, $\Psi$ and $\varphi$, similarly to what we did above. The equation for the traceless $(ij)$ component is again given by Eq.~(\ref{perturbed system ij traceless}), with $\Pi = 0$. By exploiting that equation, we easily find the Laplacian of each gravitational potential: 
\begin{equation}
        \nabla^2\Phi = -a^2\delta\varphi \frac{{\rm d}^2V}{{\rm d}\varphi^2} + \frac{W+1}{2W+3}\left(-\frac{8\pi G}{\varphi}\delta\rho + 2a^2\delta\varphi \frac{{\rm d}^2V}{{\rm d}\varphi^2}\right)\, ;
    \label{laplacian phi}
\end{equation}
 \begin{equation}
        \nabla^2\Psi = -a^2\delta\varphi \frac{{\rm d}^2V}{{\rm d}\varphi^2} + \frac{W+1}{2W+3}\left(-\frac{8\pi G}{\varphi}\delta\rho + 2a^2\delta\varphi \frac{{\rm d}^2V}{{\rm d}\varphi^2}\right) + \frac{8\pi G}{\varphi}\delta\rho
          \label{laplacian psi}\, .
    \end{equation}
    
This result shows that our choices of setting $V(\varphi)=-\Xi\varphi$ and $W(\varphi)=-1$ cause the Laplacian of the gravitational potential $\Phi$ to become zero, while the Laplacian of $\Psi$ is, as expected, proportional to the baryonic over-densities multiplied by the boost factor $\varphi^{-1}$. The lensing convergence term, which describes the change in flux due to the weak gravitational lensing, is given by~\cite{Dodelson} 
\begin{equation}
    \kappa = -\frac{1}{2}\int^\chi_0\,{\rm d}\tilde{\chi} \nabla^2\left[\Phi(\boldsymbol{x}(\boldsymbol{\theta}, \tilde{\chi}))-\Psi(\boldsymbol{x}(\boldsymbol{\theta}, \tilde{\chi}))\right]\frac{\tilde{\chi}}{\chi}(\chi-\tilde{\chi})\, ,
\end{equation}
where $\boldsymbol{x}(\boldsymbol{\theta},\tilde{\chi})$ is the unperturbed photon path, with $\boldsymbol{\theta}$ being the angular position of the lensed source and $\tilde{\chi}$ the covariant distance measured along the geodesic. Therefore, in our case, the fact that $\nabla^2\Phi$ turns out to be zero leads to a reduction of the lensing effect. This result provides additional strong constraints to the theory, suggesting that another choice of the two potentials $W(\varphi)$ and $V(\varphi)$ might be necessary.

\section{The quasi-static approximation}
\label{sec:QSA}

The results of Section \ref{growth factor equation} show that the conditions for the application of the quasi-static approximation (QSA)~\citep[][]{Silvestri2013,DeFelice2012,Pace2019,Pace2021} are valid: perturbations in the scalar field tend to become negligible and are irrelevant for the evolution of the growth factor.  In this section, we adopt the QSA and we use it to provide an alternative,  straightforward derivation of the results of Section~\ref{CRG linear perturbation theory}.

In QSA, it is assumed that the time derivatives of the potentials and of the scalar field are subdominant with respect to the spatial derivatives, as they are of the order of the Hubble-Lema\^\i tre function $H$. This condition is generally true for Horndeski models, but it is not necessarily the case for models beyond Horndeski (see, e.g.,~\cite{Lombriser2015}). QSA only works on sufficiently small scales, namely sub-horizon scales, whereas the approximation does not hold at larger scales. Specifically, QSA is valid on scales smaller than the sound horizon $c_{\rm s,\delta\varphi}k/(aH) \equiv c_{\rm s,\delta\varphi}\mathrm{K}$, with $\mathrm{K} \equiv k/(aH)$. In our case, the sound speed of the scalar field is luminal (Sect.~\ref{sec:soundspeed}), enabling us to use QSA.

The advantage of QSA is that it reduces a system of differential equations, namely the field equations and the equation of motion of the perturbed scalar field, to a system of algebraic equations. By solving for the two potentials and the scalar field, one can find three Poisson-like equations, which relate the potentials and the scalar field to the matter content of the universe, which, in our case, coincides with the baryonic matter alone. In general, the Newtonian gravitational constant $G$ is not constant, but it is a generic function of time and scale and it can be expressed in terms of the $\alpha$ functions defined in Sect.~\ref{sec:soundspeed} [see Eq.~(\ref{alphas})]. Here, we consider the two effective gravitational constants $\mu_{Y}$ and $\mu_{Z}$ and the slip $\eta = -\mu_{Z}/\mu_{Y}$. One can also define a third gravitational constant $\mu_W = \Sigma = (\mu_{Y}-\mu_{Z})/2$ that describes the modification to weak gravitational lensing. From a physical point of view, only $\mu_{Y}$ represents the effect of the modification of gravity on non-relativistic particles, whereas the effects of $\mu_{Z}$ are not directly observable, since no observable which depends on $\mu_{Z}$ alone is known. 

Here, we only report the resulting expressions for the different quantities and refer to~\cite{Pace2021} for a detailed derivation.
Following the notation of~\cite{Pace2021}, we can write the three functions as
\begin{align}
 \mu_{Z} = &\, \frac{\mu_{Z,+0} + \mu_{Z,+2}\mathrm{K}^2}{\mu_{-0} + \mu_{-2}\mathrm{K}^2}\frac{1}{\bar{M}^2}\,,\\
 \mu_{Y} = &\, \frac{\mu_{Y,+0} + \mu_{Y,+2}\mathrm{K}^2}{\mu_{-0} + \mu_{-2}\mathrm{K}^2}\frac{1}{\bar{M}^2}\,,\\
 \eta = &\, \frac{\mu_{Z,+0} + \mu_{Z,+2}\mathrm{K}^2}{\mu_{Y,+0} + \mu_{Y,+2}\mathrm{K}^2}\, .
\end{align}
The corresponding Poisson equations for the two potentials are
\begin{equation}
 \nabla^2\Psi = 4\pi G\mu_Y a^2 \delta\rho_{\rm m}\,, \quad 
 \nabla^2\Phi = -4\pi G\mu_Z a^2 \delta\rho_{\rm m}\,.
\end{equation}
The constants appearing in the three  functions above are $\bar{M}^2 \equiv M^2/M_{\text{pl}}^2$ and
\begin{align*}
 \mu_{Y,+0} & = (1+\alpha_{T})\mu_{\rm p}\,, & \mu_{-0} & = \mu_{\rm p}\,, & \mu_{Z,+0} & = \mu_{\rm p}\,, \\
 \mu_{Y,+2} & = \alpha c_{\rm s}^2\bar{M}^2\mu_{Y,\infty}\,, & \mu_{-2} & = \alpha c_{\rm s}^2\,, & \mu_{Z,+2} & = \alpha c_{\rm s}^2\bar{M}^2\mu_{Z,\infty}\,,
\end{align*}
where
\begin{equation}
 \mu_{\rm p} = 6\left\{\left(\dot{H} + \frac{\rho_{\rm m}+P_{\rm m}}{2M^2}\right)\dot{H} + \dot{H}\alpha_B\left[H^2(3+\alpha_M)+\dot{H}\right] + H\frac{\partial\left(\dot{H}\alpha_B\right)}{\partial t}\right\} \, H^{-4}\,.
\end{equation}

These general expressions cannot hold for every value of ${\rm K}^2$, because QSA requires $\rm K>1$.  
Therefore, since our Brans-Dicke theory is similar, in its structure, to a Jordan--Brans--Dicke (JBD) theory,\footnote{The Lagrangian of the JBD model is $\mathcal{L} \propto \varphi R - \omega X/\varphi - 2V(\varphi)$, which coincides with our setup if $\omega = -1$, as in dilaton gravity~\citep{Gasperini2003}.} which is known to be a scale-independent model, we can simply consider the values of  $\mu_Z$, $\mu_Y$, and $\eta$ at infinity, as the coefficient $\mu_{\rm p}$ is, in general, negligible compared to ${\rm K}^2$. Hence, the three expressions simplify to
\begin{align}
 \mu_Z=&\,\mu_{Z,\infty} = \, \frac{\alpha c_{\delta\phi}^2 + 2\alpha_B[\alpha_B(1+\alpha_T) + \alpha_T - \alpha_M]}{\alpha c_{\delta\phi}^2 \bar{M}^2}\,, \\
 \mu_Y=&\,\mu_{Y,\infty} = \, \frac{\alpha c_{\delta\phi}^2(1+\alpha_T) + 2[\alpha_B(1+\alpha_T) + \alpha_T - \alpha_M]^2}{\alpha c_{\delta\phi}^2 \bar{M}^2}\,, \\
 \eta=&\,\eta_{\infty} = \, \frac{\alpha c_{\delta\phi}^2 + 2\alpha_B[\alpha_B(1+\alpha_T) + \alpha_T - \alpha_M]}{\alpha c_{\delta\phi}^2(1+\alpha_T) + 2[\alpha_B(1+\alpha_T) + \alpha_T - \alpha_M]^2}\,.
\end{align}
By substituting the values of the different terms from Eqs.~(\ref{alphas}), we find, reminding that $\Sigma = (\mu_{Y}-\mu_{Z})/2$,
\begin{equation}
 \mu_{Z,\infty} = 0\,, \quad
 \mu_{Y,\infty} = \frac{2}{\varphi}\,, \quad
 \eta_{\infty} = 0\,, \quad
 \Sigma_{\infty} = \frac{1}{\varphi}\,.
\end{equation}

These results agree with Eqs.~(\ref{laplacian phi}) and~(\ref{laplacian psi}): 
the baryonic matter experiences stronger gravity due to the term $\mu_{Y,\infty}=2/\varphi$; at the same time, the potential $\Phi$, linked to the term $\mu_{Z,\infty}$, which is totally independent of the matter distribution (Sect. \ref{laplacian of the gravitational potentials}), leads to a null slip parameter $\eta_{\infty}$. From the parameter $\Sigma_{\infty}$, we see that, as anticipated, weak gravitational lensing is also affected, with a strength which is half the strength affecting the non-relativistic particles. Hence,
the combination of gravitational lensing information with the measures of the growth of structure in our Brans--Dicke theory is expected to provide a distinctive signature.

We conclude by illustrating why $\mu_{Z_\infty}$ is null. We consider a more generic Lagrangian, with Horndeski functions
\begin{equation}
    G_2({\varphi,X}) = -\frac{1}{2}M_{\rm pl}^2\left[f_2(\varphi)X + V(\varphi)\right]\,, \quad 
    G_4 = \frac{1}{2}M_{\rm pl}^2f_4(\varphi)\,,
\end{equation}
where $f_2(\varphi)$, $f_4(\varphi)$, and $V(\varphi)$ are arbitrary functions of the scalar field. For what follows, $V(\varphi)$ does not enter any expression, but it is important to determine the background evolution of the scalar field $\varphi$.

The $\alpha$ functions characterizing the perturbations are now
\begin{equation}
 M^2 = M_{\rm pl}^2f_4\,, \quad 
 \alpha_M = 2\alpha_B = \frac{\dot{\varphi}f_{4,\varphi}}{Hf_4}\,, \quad 
 \alpha_K = \frac{f_2\dot{\varphi}^2}{H^2f_4}\,,
\end{equation}
where we did not explicitly write the dependence on the scalar field. Finally, we consider the quantity
\begin{equation}
 \rho_{\varphi} + P_{\varphi} = M_{\rm pl}^2\left[\left(f_2+f_{4,\varphi\varphi}\right)\dot{\varphi}^2 + \left(\ddot{\varphi}-H\dot{\varphi}\right)f_{4,\varphi}\right]\,.   
\end{equation}
Inserting these equations into the expression for the sound speed, we find $c_{\rm s,\delta\phi}^2 = 1$. Using this result in the expression for $\mu_{Z,\infty}$, this quantity is null only if
\begin{equation}
    \alpha_K + 4\alpha_B^2 = 0 \rightarrow f_2f_4+f_{4,\varphi}^2 = 0\,.
\end{equation}
Hence, the following relation between $f_2$ and $f_4$ must hold
\begin{equation}
    f_2 = -\frac{f_{4,\varphi}^2}{f_4}\,.
\end{equation}
If we now set $f_{4,\varphi} = \varphi$, as in our Brans-Dicke theory, we obtain $f_2(\varphi) = -1/\varphi$, which coincides with our expression in Eq.~(\ref{Horndeski functions of CRG}) for $G_2(\varphi,X)$ if $W(\varphi) = -1$. Thus, it is our particular choice of the coupling $W(\varphi)$ and a linear $V(\varphi)$, which leads to $\mu_{Z,\infty} = 0$, as anticipated above by Eq.~(\ref{laplacian phi}).

\section{Conclusions}
\label{Conclusions}

We investigate the formation of the large-scale cosmic structures in a universe with baryonic matter alone. Gravity is governed by a scalar-tensor theory of gravity belonging to the class of the Brans--Dicke theories. In the weak-field limit, this theory reduces to RG, that describes the dynamics of galaxies and galaxy clusters without the need for dark matter~\cite{Matsakos, Cesare:2020pyz, Cesare:2022khd, Pizzuti}. The theory is specified by the two functions of the scalar field $W(\varphi)=-1$ and $V(\varphi)=-\Xi\varphi$, with $\Xi$ an arbitrary positive constant. 

The accelerated expansion of the homogeneous universe is driven by the scalar field $\varphi$. On small scales, the same scalar field generates the gravitational boost that plays the role of dark matter in the standard model. Therefore, in this theory, both sides of the dark sector of the standard model is replaced by a single scalar field. Unlike other UDM models~\cite{Bertacca:2008uf, Bertacca:2010ct}, however, in our Brans--Dicke theory the scalar field is not expected to cluster: indeed, we show that the sound speed of the scalar field coincides with the speed of light.

We solve the modified Friedmann equations for a homogeneous universe with flat geometry and derive the redshift evolution of the Hubble-Lema\^\i tre parameter $H(z)$. The comparison of the expected evolution of $H(z)$ with measures based on cosmic chronometer data sets, up to $z\sim 2$, returns the free cosmological parameters. 
In agreement with previous results based on high-redshift type Ia supernovae~\cite{Sanna}, the comparison suggests $\Omega_{\Xi 0}=\Xi/3H_0^2\sim 0.5$ and $\Omega_0=2\Omega_{\mathrm{m} 0}/\varphi_0\sim 0$, where the index $0$ refers to the quantities at the present time. 

We derive the equation of the growth factor in linear perturbation theory. The solution to this equation shows that large-scale structure do form, but the non-linear regime is reached at redshift $z< 1$, at odds with the observation of massive galaxies at redshift $z\gtrsim 10$~\cite{naidu2025cosmicmiracleremarkablyluminous}. We show that this delayed formation is generated by the coefficient of the source term in the growth factor equation. This coefficient is proportional to $1/[\varphi(a) H^2(a) a^3]$, with $a$ the scale factor. The choice $V(\varphi)\propto \varphi$ returns the background scalar field evolution $\varphi(a)\propto H^{-1}(a)a^{-3}$ in a homogeneous universe. The coefficient in the growth factor equation thus scales as $H^{-1}(a)$ and the source of the gravitational field is relevant only at low redshift.

We derive the Laplacians of the two gravitational potentials $\Phi$ and $\Psi$ that appear in the perturbed FLRW metric. The potential $\Phi$ has null Laplacian, whereas the Laplacian of $\Psi$ is proportional to the matter density times the factor $2/\varphi$. It follows that the gravity acting on massive particles, which is proportional to $\Psi$, is twice the gravity acting on photons, which is proportional to $(\Phi+\Psi)/2$. In other words, the dynamics of massive particles and gravitational lensing are governed by a different intensity of gravity. This result derives from our choice $W(\varphi)=-1$ and a linear $V(\varphi)$ and, in principle, should have distinctive observational consequences.

We finally show that the growth factor equation satisfies the conditions of the quasi-static approximation. The adoption of this approximation confirms the difference of the gravity intensity acting on massive particles and on photons.

In conclusion, our analysis shows that the covariant extension of RG in the form of this Brans--Dicke theory fails to account for the observed timing and amplitude of structure formation. It remains to be seen whether a different choice of $W(\varphi)$ and $V(\varphi)$, or a more general theory, belonging to the larger family of Horndeski theories, might solve these problems and still reduce, in the weak-field limit, to RG, which appears to be viable on the scale of galaxies and galaxy clusters.

\acknowledgments
We acknowledge partial support from the grant InDark of the Italian National Institute of Nuclear Physics (INFN).
APS is supported by the MUR FIS2 Advanced Grant ET-NOW (CUP:~B53C25001080001) and by the INFN TEONGRAV initiative. He also gratefully acknowledges support from a research grant funded under the INFN–ASPAL agreement as part of the Einstein Telescope training program.

This research has made use of NASA’s Astrophysics Data System Bibliographic Services. 

\clearpage
\appendix
\section{Ricci tensor components for the metric of Eq.~(\ref{FRW})}
\label{AppA}
In this Appendix, we provide some useful results to reproduce the equations in Section~\ref{Homogenous and isotropic universe in CRG}. The relevant components of the Ricci tensor and the Ricci scalar read
\begin{equation}
\begin{gathered}
    R_{00} = -3\frac{\ddot{a}}{a}; \hspace{3mm}    R_{rr} = \frac{1}{1-kr^2}\left(a\ddot{a} + 2\dot{a}^2 + 2k\right)\,; \hspace{3mm}
    R = 6\frac{\ddot{a}}{a} + 6\left(\frac{\dot{a}}{a}\right)^2 + 6\frac{k}{a^2}\,.
    \label{R components}
\end{gathered}
\end{equation}
Other useful quantities to evaluate the KG equation for the scalar field are
\begin{equation}
\begin{gathered}
\nabla^\alpha\varphi\nabla_\alpha\varphi = -\dot{\varphi}^2\,; \hspace{3mm}
    \Box\varphi = - \ddot{\varphi} - 3\frac{\dot{a}}{a}\dot{\varphi}\,; \hspace{3mm}
    \nabla_0\nabla_0\varphi = \ddot{\varphi}\,;\\
    \nabla_r\nabla_r\varphi = -\frac{1}{1-kr^2}a\dot{a}\dot{\varphi}\,; \hspace{3mm}
    \nabla_0\varphi\nabla_0\varphi = \dot{\varphi}^2\,.
\end{gathered}
\label{Phi derivatives}
\end{equation}

\section{Tensor perturbations}
\label{TensorPerturbations}
The derivation of the equations for the tensor perturbations follows the same steps of the scalar perturbations, but with a perturbed metric of the form
\begin{equation}
    \mathrm{d} s^{2}=a^{2}(\eta)\left[- \mathrm{d} \eta^{2} + \left(\delta_{i j}+h_{ij}^{TT}\right) \mathrm{d} x^{i} \mathrm{d} x^{j}\right]\,,
\end{equation}
where $h_{ij}^{TT}$ is a traceless and transverse $3\times3$ matrix, whose inverse satisfies $\delta_{ij}h^{TT,jk} = -h_{ij}^{TT}\delta^{jk}$. By computing the resulting Christoffel symbols, we arrive at the modified Einstein equations
\begin{equation}
    \varphi\left(-\frac{1}{2}h_{ij}^{\prime\prime} - \mathcal{H}h_{ij}^\prime + \frac{1}{2}\nabla^2h_{ij}\right) - \frac{1}{2}\varphi^\prime h_{ij}^\prime = -8\pi Ga^2\left(P\delta_{ij} + \pi_{ij}\right)\,.
\end{equation}
For a gravitational wave traveling along the $z$-direction, we can write
\begin{equation}
    h_{ij}^{TT} = \left( \begin{array}{ccc}
         h_+ & h_\times & 0 \\
         h_\times & -h_+ & 0 \\
         0 & 0 & 0
    \end{array} \right) \, , \hspace{5mm} 
    h^{TT,ij} = \left( \begin{array}{ccc}
         -h_+ & -h_\times & 0 \\
         -h_\times & h_+ & 0 \\
         0 & 0 & 0
    \end{array} \right)\, .
\end{equation}
Thus, we have two equations of motion for the two modes $h_+$ and $h_\times$:
\begin{align}
    \varphi\left(h_{+}^{\prime\prime} + 2\mathcal{H}h_{+}^\prime -\nabla^2h_{+}\right) + \varphi^\prime h_{+}^\prime = &\, 8\pi Ga^2\left(\pi_{11} - \pi_{22}\right)\,,\\
    \varphi\left(h_{\times}^{\prime\prime} + 2\mathcal{H}h_{\times}^\prime -\nabla^2h_{\times}\right) + \varphi^\prime h_{\times}^\prime = &\, 8\pi Ga^2\left(2\pi_{12}\right)\,.
\end{align}

\bibliography{biblio}

\end{document}